\pdfoutput=1
\documentclass[11pt]{article}
\usepackage{jheppub}
\usepackage{color}
\usepackage[all]{xy}
\usepackage{customprelude}

\usepackage{tabu}
\usepackage{blkarray}
\title{\centering Higher Form Symmetries and M-theory}

\author[\sharp]{Federica Albertini,}
\author[\sharp \ast]{Michele Del Zotto,}
\author[\sharp]{Iñaki García Etxebarria,}
\author[\sharp]{Saghar S. Hosseini}

\affiliation[\sharp]{Department of Mathematical Sciences\\
Durham University, Durham, DH1 3LE, United Kingdom}
\affiliation[\ast]{Department of Mathematics, and Department of Physics and Astronomy\\ Uppsala University, Uppsala, Sweden}

\emailAdd{federica.albertini@durham.ac.uk}
\emailAdd{michele.delzotto@math.uu.se}
\emailAdd{inaki.garcia-etxebarria@durham.ac.uk}
\emailAdd{sagharsadat.hosseinisemnani@durham.ac.uk}

\abstract{We discuss the geometric origin of discrete higher form symmetries of quantum field theories in terms of defect groups from geometric engineering in M-theory. The flux non-commutativity in M-theory gives rise to  (mixed) 't Hooft anomalies for the defect group which constrains the corresponding global structures of the associated quantum fields. We analyze the example of 4d $\cN=1$ SYM gauge theory in four dimensions, and we reproduce the well-known classification of global structures from reading between its lines. We extend this analysis to the case of 7d $\cN=1$ SYM theory, where we recover it from a mixed 't Hooft anomaly among the electric 1-form center symmetry and the magnetic 4-form center symmetry in the defect group. The case of five-dimensional SCFTs from M-theory on toric singularities is discussed in detail. In that context we determine the corresponding 1-form and 2-form defect groups and we explain how to determine the corresponding mixed 't Hooft anomalies from flux non-commutativity. Several predictions for non-conventional 5d SCFTs are obtained. The matching of discrete higher-form symmetries and their anomalies provides an interesting consistency check for 5d dualities.}

\newcommand\be{\begin{equation}}
\newcommand\ee{\end{equation}}

\newcommand\sing{{{\mathcal V}_6}}
\newcommand\TX{{\mathcal T}_\sing}
\newcommand\KM{{\mathbb E}_M}

\begin{document}

% \makeatletter
% \let\old@fpheader\@fpheader
% \renewcommand{\@fpheader}{\old@fpheader\hfill
% PREPRINT}
% \makeatother

\maketitle

\section{Introduction}

In this paper we are interested in the study of discrete higher form symmetries for quantum field theories that arise by geometric engineering in M-theory. In full generality, $p$ branes wrapping non-compact $k$-cycles in the non-compact internal geometry  $\mathcal V$ of geometric engineering give rise to $p-k+1$ dimensional defects for the field theory. Similarly, $p$ branes wrapping compact $k$-cycles give rise to $p-k+1$ BPS degrees of freedom. Generalizing the 't Hooft screening argument along the lines of \cite{DelZotto:2015isa},  for every $p$ brane wrapping a non-compact $k$-cycle we expect to find a corresponding defect group of discrete higher $(p-k+1)$-form symmetries,\footnote{See \cite{Kapustin:2014gua,Gaiotto:2014kfa,Sharpe:2015mja} for a definition of higher form symmetries. In this paper we adopt the C{\'o}rdova-Dumitrescu-Intriligator notation: $G^{(m)}$ denotes an $m$-form symmetry group $G$ \cite{Cordova:2018cvg}. We stress that in principle one could consider to include the non-torsional part of the groups in the definition \eqref{eq:defecto}. We choose not to include it to avoid cluttering notation below. Moreover, there might be other global symmetries arising from isometries which the defect group will not detect --- we thank Kantaro Ohomori for this remark.} 
\be\label{eq:defecto}
\mathbb D \df \bigoplus_{n} \mathbb D^{(n)} \qquad\text{where}\quad  \mathbb D^{(n)}\df \bigoplus_{p \text{ branes and } k\text{ cycles} \atop \text{such that } p-k+1 = n} \Tor \left({H_k(\mathcal V,\partial \mathcal V) \over H_k(\mathcal V)}\right)
\ee
The operators that are measuring these charges are the flux operators in the string theory we are adopting for the engineering that are sourced by the corresponding kind of $p$ brane. These operators are the charge operators for the $\mathbb D^{(p-k+1)}$ factor of the defect group. We stress that $\mathbb D$ is \textit{not} the group of higher form symmetries of a quantum field theory yet, rather it is the group of higher form symmetries acting on the \textit{geometric engineering Hilbert space}, which is the Hilbert space that the given string theory assigns to the given non-compact geometry.

\medskip

The geometric engineering formalism indeed has a deeper insight on the structure of the resulting field theories: whenever the internal manifold has torsional cycles, we have non-commuting fluxes (a fact that has been thoroughly discussed in the seminal papers by Freed, Moore and Segal \cite{Freed:2006ya,Freed:2006yc} --- see also \cite{Garcia-Etxebarria:2019cnb}). Correspondingly, pairs of electromagnetically dual branes in the geometric engineering give rise to mixed 't Hooft anomalies among different higher form symmetry factors of the defect group.

\medskip

Examples including the computation of such mixed 't Hooft anomalies in the context of the 6d 2-form symmetry defect group appeared in \cite{Witten:2009at,DelZotto:2015isa,Heckman:2017uxe,Garcia-Etxebarria:2019cnb}.\footnote{See also \cite{Gukov:2018iiq,Eckhard:2019jgg,Dabholkar:2020fde} for some deep implications.} The purpose of the present paper is to extend this study in the context of models geometrically engineered within M-theory. Our main aim is to explain how the discrete higher form symmetries arise in this context and what is their relation with the M-theory defect group. It is well-known that geometric engineering in string theory gives an alternative formulation of field theories that often proves useful when studying models that cannot be realized perturbatively, which is the case for all SCFTs in dimension higher than four  \cite{Cordova:2016emh,Chang:2018xmx}. Among the most interesting examples in this class therefore are 5d SCFTs, whose geometric engineering  \cite{Seiberg:1996bd,Morrison:1996xf,Intriligator:1997pq,Aharony:1997ju,Leung:1997tw} has seen a lot of recent developments (for an incomplete list of recent references see \cite{Bergman:2013ala,Hayashi:2014hfa,Bergman:2014kza,Zafrir:2014ywa,DelZotto:2014hpa,Bergman:2015dpa,Zafrir:2015ftn,Kim:2016qqs,DelZotto:2015rca,Hayashi:2015fsa,Hayashi:2015vhy,Hayashi:2016abm,Xie:2017pfl,DelZotto:2017pti,Alexandrov:2017mgi,Ferlito:2017xdq,Jefferson:2017ahm,Jefferson:2018irk,Apruzzi:2018nre,Bhardwaj:2018yhy,Closset:2018bjz,DelZotto:2018tcj,Bhardwaj:2018vuu,Bhardwaj:2019hhd,Bhardwaj:2019ngx,Bhardwaj:2019fzv,Bhardwaj:2019xeg,Bhardwaj:2020gyu,Bhardwaj:2020kim,Apruzzi:2019kgb,Apruzzi:2019enx,Apruzzi:2019vpe,Apruzzi:2019opn,Closset:2019juk,Kim:2019dqn,BenettiGenolini:2019zth,Hayashi:2019fsa,Bourget:2020gzi,Bourget:2019rtl,Cabrera:2019izd,Cabrera:2018jxt,Hayashi:2019jvx,Hayashi:2019yxj,Hayashi:2018lyv,Hayashi:2017jze,Hayashi:2015zka,Saxena:2019wuy,Garozzo:2020pmz}). Among the applications of our formalism, we determine the global structure and the discrete higher form symmetries of 5d SCFTs. A more detailed summary of our results can be found in \S \ref{sec:summary} below.

\subsection{Discrete higher form symmetries from M-theory: General philosophy}

In this work we consider the SQFTs obtained from M-theory via geometric engineering on backgrounds of the form
\be\label{eq:ilgeneralesullacollina}
\cM_{11} = \cM_{D} \times \mathcal V_{d} \qquad d+D = 11
\ee
where $\mathcal V_{d}$ is a local internal geometry for M-theory and $\mathcal M_{D} $ is a $D$ dimensional space time manifold where the geometrically engineered $D$ dimensional quantum field theory $\mathcal T_{\mathcal V_d}$ lives. 

\medskip

\noindent For simplicity, we assume:\footnote{Both assumptions can be dropped in principle, but we want to work in the simplest possible setup that highlights the features we want to study.}
\begin{itemize}
\item $\mathcal V_{d}$ is a supersymmetric background, therefore $\mathcal T_{\mathcal V_d} \in \text{SQFT}_{D}$; moreover
\be
\mathcal V_d = \mathcal C(Y_{d-1})\,,
\ee
meaning that $\mathcal V_d$ is a metric cone over a $d-1$ dimensional manifold $Y_{d-1}$;
\item $\cM_{D}$ is a closed spin manifold without torsion.
\end{itemize}
Naively, this M-theory setup computes the partition function of the theory $\mathcal T_{\mathcal V_d}$ on the manifold $\cM_{D}$,
\be
Z_{\mathcal T_{\mathcal V_d}} (\cM_{D}) \in \bC\,.
\ee
The latter should be fully specified by the M-theory background, but in presence of mixed 't Hooft anomalies for the factors in the defect group not all fluxes can be diagonalized simultaneously, thus leading to several different choices. These choices are in one-to-one correspondence with the possible global structures of the quantum field theory $\mathcal T_{\mathcal V_d}$.
\bigskip

The main feature of the geometric engineering limit is that the internal manifold is non-compact, and therefore the M-theory background has a boundary at infinity
\be
\partial \mathcal \cM_{11} = \cM_D \times \partial \mathcal V_d\,.
\ee
If this is the case, we can consider a Hamiltonian quantization viewing the direction normal to the boundary as time, and assign to this system a Hilbert space 
\be
\mathcal H_M(\partial \mathcal \cM_{11} )\,.
\ee
By analogy with the IIA and IIB superstrings, the Hilbert space $\mathcal H_M(\partial \mathcal \cM_{11})$ should have a grading in terms of the \textit{M-theory generalized cohomology theory group}, which we denote $\KM(\partial \mathcal \cM_{11})$. The group $\KM(\partial \mathcal \cM_{11})$ is expected to parametrize the flux sectors of M-theory. Classically one would expect all such fluxes to commute, but this is not the case because the group $\KM(\partial \mathcal \cM_{11})$ can contain a torsional part 
\be
\text{Tor } \KM(\partial \mathcal \cM_{11}) = \{ x \in \KM(\partial \mathcal \cM_{11}) \colon n x = 0 \text{ for some } n \in \mathbb Z\}\,
\ee
Under the assumptions above, there is a simple connection in between the defect group and $\Tor \KM (\partial \cM_{11})$: 
\be\label{eq:themagic}
\Tor \KM (\partial \cM_{11}) = \bigoplus_{j} H^{j+1}(\cM_{D}) \otimes \mathbb D^{(j)}
\ee
In presence of torsional fluxes we expect to have a grading of the geometric engineering Hilbert space in terms of
\be
\mathcal H_M = \bigoplus_{\boldsymbol{\alpha} \in\KM^o(\partial \mathcal \cM_{11})} \mathcal H_M (\boldsymbol{\alpha}) \qquad\text{with}\quad \KM^o(\partial \mathcal \cM_{11}) \equiv {\KM(\partial \mathcal \cM_{11}) \over \text{Tor } \KM(\partial \mathcal \cM_{11})}
\ee
where each factor $\mathcal H_M (\boldsymbol{\alpha})$ is in turn a representation of a Heisenberg algebra of torsional fluxes of the form
\be\label{eq:thegroup}
\Psi_x \Psi_y = s(x,y) \Psi_y \Psi_x
\ee
where
\be
s \colon \text{Tor } \KM(\partial \mathcal \cM_{11}) \times \text{Tor } \KM(\partial \mathcal \cM_{11}) \to U(1)
\ee
is a perfect pairing. \textit{This pairing encodes the mixed 't Hooft anomalies among the higher form symmetries of the geometric engineering Hilbert space.} Abusing language, in light of \eqref{eq:themagic} we will refer to these as \textit{the mixed 't Hooft anomalies for the defect group}. Since the flux operators do not commute we cannot specify the asymptotic values for all fluxes simultaneously: two steps are required
\begin{enumerate}
\item We need to choose a maximally isotropic subgroup $L \subset \text{Tor } \KM(\partial \mathcal \cM_{11})$ of fluxes that can be simultaneously measured;
\item We need to choose a ``zero flux'' state, which corresponds to the unit eigenvalue
\be
\Psi_x |0,L\rangle = |0,L\rangle \qquad \forall x \in L
\ee 
\end{enumerate}
\noindent Then we obtain a basis for the geometric engineering Hilbert space parametrized by
\be
|f,L\rangle \df \Psi_f |0,L\rangle \qquad\quad f \in \sF_L \df {\text{Tor } \KM(\partial \mathcal \cM_{11}) \over L}\,.
\ee
A choice of background fluxes for the higher form symmetries of this theory corresponds to fixing a state
\be
|\{a_f\}\rangle = \sum_{f\in \sF_L} a_f |f,L\rangle\,
\ee
whence the corresponding partition function is determined: the open manifold $\cM_{11}$ can be viewed as an element $\langle \cM_{11} | $ of $\mathcal H_M(\partial \mathcal \cM_{11})^* \df \Hom (\mathcal H_M(\partial \mathcal \cM_{11}),\mathbb C)$ so the partition function is $\langle \cM_{11} |\{a_f\}\rangle$ --- see \cite{Garcia-Etxebarria:2019cnb} for a more detailed version of this argument.

\medskip

We stress that to fully specify a quantum field theory $\mathcal T_{\mathcal V_d}$ these two steps are required. Indeed only in this case we end up with a partition function. Without specifying these details, the geometric engineering Hilbert space knows only about the whole set of possible theories that have the same local dynamics but different global structures. When we choose the theory corresponding to the state $|0,L\rangle$, the defects with charges in $L$ are non-genuine, while the ones with charges in $\sF_L$ are the genuine ones (see section 3.3 of \cite{Garcia-Etxebarria:2019cnb}). When we specify the state $|0,L\rangle$ this breaks the defect group $\mathbb D$ to the higher form symmetry group of the genuine defects of the corresponding quantum field theory $\mathcal T_{\mathcal V_d}$. The operators $\Psi_f$ generate background flux for the higher form symmetry associated to the genuine operators. We can think of them as domain walls and of the states $|f,L\rangle$ as labeling distinct superselection sectors.

\medskip

To clarify this statement let's consider the example of geometric engineering of 4d $\cN=1$ SYM with simple simply-laced Lie algebra $\fg_\Gamma$ where $\Gamma \subset SU(2)$. The defect group we obtain for this geometry is
\be
\mathbb D = Z(G_\Gamma)^{(1)}_{M2} \oplus Z(G_\Gamma)^{(1)}_{M5}\,.
\ee
We have an electric and a magnetic one form symmetry valued in the center of $G_\Gamma$, the universal cover group. These one-form symmetries of the geometric engineering Hilbert space however have a mixed 't Hooft anomaly: the corresponding charge operators do not commute. More precisely, for fluxes labelled by $a_i = (\omega \otimes \ell)_i \in H^2(\cM_4)\otimes Z(G_\Gamma)$ we have
\be
\Phi_{M2,a_1} \Psi_{M5,a_2} = \exp \left( 2\pi i  \, \mathsf{L}_\Gamma(\ell_1,\ell_2) \int_{\cM_4} \omega_1 \wedge \omega_2\right) \Psi_{M5,a_2} \Phi_{M2,a_1}
\ee 
where $\mathsf{L}_\Gamma(\ell_1,\ell_2)$ is a perfect pairing in $Z(G_\Gamma)$ that is determined below --- see table \ref{ADE-linking-forms}. Specifying a state now selects a surviving subgroup of $\mathbb D$ which become the 1-form symmetry for the SYM quantum field theory. Consider for example the case $\Gamma = \mathbb Z_N$. For instance, we can choose a maximal isotropic lattice $L_{M2}$ generated by the M2 flux operators. Then the state $|0,L_{M2}\rangle$ corresponds to the theory $PSU(N) = SU(N)/\mathbb Z_N$ and the wrapped $M5$ branes are the genuine defects charged under the resulting magnetic 1-form symmetry.  Conversely choosing to set to zero all M5 fluxes we are selecting the state $|0,L_{M5}\rangle$: we are preserving the electric 1-form symmetry, thus leading to the theory with gauge group $SU(N)$.

\subsection{Summary of results and structure of this work}\label{sec:summary}

The structure of this work is as follows. As a simpler warm-up example, in section \ref{sec:7d} we discuss the case of 7d gauge theories with simple simply-laced lie groups $G\in ADE$; of course, we find agreement with the global structure obtained by considering Wilson and 't Hooft operators in the 7d side, and the global structure predicted by M-theory flux non-commutativity. In section \ref{sec:5d-general} we set the stage for our analysis of the global structure of 5d SCFTs from M-theory on canonical CY singularities. In this work, for simplicity, we consider geometries with some constraints on the cohomology which restrict the structure of the models we analyze. As a result we obtain 5d defect groups of the form\footnote{ In principle one could expect to obtain a further factor $\tilde{\sZ}_{M2}^{(2)} \oplus \tilde{\sZ}_{M5}^{(1)}$, for geometries with nontrivial $H_1(\mathcal V_6)$ and $H_5(\mathcal V_6)$ that we assume to vanish for simplicity in chasing exact sequences.}
\be
\mathbb D = \Big(\sZ^{(1)}_{M2} \oplus \sZ^{(2)}_{M5}\Big) \oplus \Big(\sZ^{(-1)}_{M2} \oplus \sZ^{(4)}_{M5}\Big) 
\ee
that often present very interesting global structures arising from the mechanism we have outlined above. Here $\sZ$ is an abelian discrete group given by the torsional part of the cokernel of the intersection matrix of the corresponding CY. Physically this intersection matrix is associated to the Dirac pairing among the monopole strings and the BPS particles of the SCFT in a Coulomb phase and $\sZ$ measures the 't Hooft charges of the defects. We begin reviewing some field theory results, and then proceed determining the corresponding defect groups from geometry first in general, and then focussing on the case of toric canonical CY singularities, that are the main class of examples we use for this project. In section \ref{sec:su(p)_q} the case of the 5d Yang-Mills theories with gauge algebra $\mathfrak{su}(p)$ and Chern-Simons level $k$ is studied in detail, as a consistency check for our methods. In section \ref{sec:5duality} we discuss applications of the higher form symmetries to the study of 5d dualities among different gauge theory phases of 5d SCFTs. Exploiting consistency with dualities we extend (and prove) a new purely graphical prescription to compute the defect groups for arbitrary toric CY singularities. We proceed determining the defect groups for several examples of 5d SCFTs with non-trivial flavor symmetries, corresponding to non-isolated singularities, that we study in section \ref{sec:nonlagra}. An interesting class of examples that we consider is given by the higher rank generalizations of the $E_0$ 5d SCFT. We find that the corresponding geometry gives a defect group
\be
\mathbb D = (\mathbb Z_{2r+1})^{(1)}_{M2} \oplus(\mathbb Z_{2r+1})^{(2)}_{M5} 
\ee
with a nontrivial global structure with pairing $1/(2r+1)$. From this result it follows that there are as many inequivalent versions of the 5d $E_0^{(r)}$ theories as there are inequivalent versions of $\mathfrak{su}(2r+1)$ in 4d. These models enjoy different combinations of the electric/magnetic 1-form and 2-form symmetries in the defect group. In particular we have a purely electric 5d $E_0^{(r)}$ theory which has higher form symmetry $\mathbb Z_{2r+1}^{(1)}$ and a purely magnetic 5d $E_0^{(r)}$ theory that has higher form symmetry $\mathbb Z_{2r+1}^{(2)}$. These models have identical spectra of local operators, but different properties which can be detected from the spectrum of nonlocal operators. Using our results we show that  the 5d trinions (the $T_N$ theories) have trivial defect groups in agreement with a result by Tachikawa \cite{Tachikawa:2013hya}. However, inspecting theories that can be obtained by massive deformations of the $T_N$ theories it is easy to find models that have much larger defect groups, for instance we find a family of models that have
\be
\mathbb D = (\mathbb Z_{N^2-3(N-1)})^{(1)}_{M2} \oplus(\mathbb Z_{N^2-3(N-1)})^{(2)}_{M5} 
\ee
We conclude the discussion with some applications of 5d dualities and higher form symmetries to constrain the properties of the spectrum of operators of several strongly interacting 5d SCFTs that do not admit gauge theory phases. In section \ref{sec:4d} we discuss some appetizer about the case of 4d $\cN=1$ theories arising from $G_2$ spaces. We find that the defect group of 4d $\cN=1$ models from M-theory has the structure
\be
\mathbb D = \Big(\sZ^{(1)}_{M2} \oplus \sZ^{(1)}_{M5}\Big)\oplus \Big(\tilde{\sZ}^{(0)}_{M2} \oplus \tilde{\sZ}^{(2)}_{M5}\Big) \oplus \Big(\tilde{\sZ}^{(-1)}_{M2} \oplus \tilde{\sZ}^{(3)}_{M5}\Big)
\ee
Exploiting this result we correctly reproduce the global structure of 4d $\cN=1$ pure SYM theories with gauge algebra $\fg$ from geometry --- these models have $\tilde \sZ = 0$ and $\sZ = Z(G_\Gamma)$ above. In general, however, 4d $\cN=1$ theories admit a richer global structure consisting of two commuting Heisenberg algebras ($\sZ \neq \tilde{\sZ}$ above) we also discuss some of the implications of this remark.

\bigskip

\textit{Note added: while this work was in preparation, we learned about \cite{Morrison:2020ool} that overlaps with some of our results. We thank the authors of that paper for agreeing to coordinate the submission with us.}

\section{Global structures of $\cN=1$ seven-dimensional theories from M-theory}\label{sec:7d}

We start with the case of seven dimensional $\cN=1$ theory with gauge
algebra $\fg_\Gamma$, with $\Gamma\subset SU(2)$ an ADE
group.\footnote{It is possible to consider non-simply-laced cases too
  in M-theory using frozen fluxes \cite{deBoer:2001wca}. We assume
  that no such fluxes are present, but it would be certainly be
  interesting to understand how the discussion gets modified in this
  case.}  Such theories can be engineered by considering M-theory on $\cM_{11}=\cM_7 \times \bC^2/\Gamma$, with $\Gamma$ a discrete subgroup of $SU(2)$.\footnote{ Recent results about the partition functions of seven-dimensional
gauge theories on curved spaces \cite{Minahan:2015jta,Polydorou:2017jha,Iakovidis:2020znp} should have an M-theory interpretation in terms of the M-theory Hilbert space $\cH_M(\partial \cM_{11})$ associated to this background.}

\medskip From the field theory side, we expect to have a one-form
symmetry measuring Wilson lines, and a $7-3=4$-form symmetry measuring
't Hooft surfaces (which are operators wrapping 4-surfaces in seven
dimensions). Or slightly more generally, we have electric charge
operators of dimension $7-1-1=5$, associated to elements of
$H^2(\cM_7)$, measuring the flux that would be created by Wilson
lines, and magnetic charge operators of dimension $7-4-1=2$,
associated to elements of $H^5(\cM_7)$, measuring the flux created by
't Hooft surfaces. It is useful to make this distinction,
since on manifolds of non-trivial topology it is possible to introduce
the fluxes without introducing the extended operators
themselves. These extended charge operators are valued on
$Z(G_\Gamma)$, the centre of the universal cover of any gauge group
with algebra $\fg_\Gamma$. For the ADE cases we have
$Z(G_\Gamma)=\Gamma^{\text{ab}}\df [\Gamma, \Gamma]$, the
abelianization of $\Gamma$ (see table~\ref{ADE-linking-forms} below).

Not all such higher form symmetries are present in any given theory
simultaneously, though: since the Wilson line operators are not
mutually local with respect to the 't Hooft surfaces it is not
possible to construct charge operators measuring all such charges
at the same time. What we can do instead is --- as in
\cite{Gaiotto:2010be,Aharony:2013hda} --- to choose a maximal set of mutually local
Wilson/'t Hooft operators, and declare that these are the genuine
ones.

We refer to the choice of $p$-form charge operators present in the
theory as a choice of global form for the theory and an actual choice
of flux for these operators as a background for the $p$-form symmetry
in that theory. If we sum over fluxes in
$H^2(\cM_7;\Gamma^{\text{ab}})_m$ we would have the $G_\Gamma/\Gamma$
theory ($SU(N)/\bZ_N$, for instance, in this context the fluxes are
often known as the generalized Stiefel-Whitney classes of the bundle),
with a 4-form symmetry, while if we sum over
$H^5(\cM_7;\Gamma^{\text{ab}})_e$ instead we have the $G_\Gamma$
theory ($SU(N)$, for instance) with a 1-form symmetry. We emphasize
that in a purely perturbative presentation the notion of ``sum over
fluxes in $H^5(\cM_7;\Gamma^{\text{ab}})_e$'' is somewhat formal, as
there are no fields in the Lagrangian that can detect these
fluxes. Nevertheless this choice of language becomes very natural from
the string theory point of view (and also for four dimensional
theories with electromagnetic duality, although we will not consider
such examples in this paper), so we will still adopt it.

\medskip

Let us now discuss how to reproduce these results from the M-theory
perspective, along the lines of \cite{Garcia-Etxebarria:2019cnb}. The
key fact is that in the presence of torsion at infinity the boundary
values for the $F_4$ and $F_7$ fluxes do not commute
\cite{Freed:2006ya,Freed:2006yc,Garcia-Etxebarria:2019cnb}. The spaces
in which we are engineering the seven dimensional $\fg_\Gamma$ theory
are non-compact, so strictly speaking they have no boundary, but we
will assume that whenever we have a space of the form
$\cM_{11} = \cM_{p}\times \cC(\cN_{10-p})$, with $\cC(\cN_{10-p})$
asymptotically a cone over $\cN_{10-p}$, then the right prescription
for choosing asymptotic values for the fields is to quantize the
theory on $\bR_t\times \cM_{10}$, with
$\cM_{10}\df \cM_p\times \cN_{10-p}$, and to choose a state in the
Hilbert space associated to $\cM_{10}$. This prescription has appeared
previously in the context of AdS/CFT
\cite{Aharony:1998qu,Witten:1998wy}, and it was shown in
\cite{Garcia-Etxebarria:2019cnb} to lead to the right predictions in
the context of six-dimensional SCFTs and $\cN=4$ theories in four
dimensions.

\medskip

We are (thankfully) only interested in the grading of the Hilbert
space of M-theory by topological class of the flux. In general,
M-theory fluxes live on some cohomology theory $\KM$, which is known
not to be ordinary (singular, say) cohomology. A rather dramatic
effect that can take us away from ordinary cohomology is that there is
a shifted flux quantization condition \cite{Witten:1996md}:
\begin{equation}
  \label{eq:G_4-shifted-quantization}
  \left[\frac{G_4}{2\pi}\right] - \frac{p_1(\cM_{11})}{4} \in H^4(\cM_{11};\bZ)
\end{equation}
with $p_1(\cM_{11})$ the first Pontryagin class of the tangent bundle
of $\cM_{11}$. Fortunately this shifted quantization condition will
not affect our discussion in any significant way, since
$p_1(\cM_p\times \cC(\cN_{10-p}))=p_1(\cM_p)+p_1(\cC(\cN_{10-p}))$,
which has legs either purely along $\cM_p$ or $\cC(\cN_{10-p})$. The
fluxes of interest to us, on the other hand, have legs along both
components. (An exception to this statement are fluxes associated with
$(-1)$-form symmetries that we will encounter below, but in this paper
we will not try to understand these in any detail.) Due to this fact
we will use ordinary singular cohomology in our calculations
below.\footnote{The shifted quantization
  condition~\eqref{eq:G_4-shifted-quantization} is not necessarily the
  only issue with using singular cohomology, see for instance
  \cite{Sati:2013rxa} for a discussion of further subtleties, and a
  proposal for a generalized cohomology theory taking them into
  account. The good agreement between the results from singular
  cohomology and field theory in the examples below encourages us to
  think that, at least in the simple backgrounds that we consider, our
  assumption for $\KM$ is correct, but it would certainly be
  interesting to try and find sufficiently complicated backgrounds
  where singular cohomology is not enough, and understand the field
  theory implications of this fact.}

\medskip

The screening argument we discussed in the introduction gives the following defect group for the geometric engineering Hilbert space of this theory
\be
\mathbb D = Z(G_\Gamma)^{(1)}_{M2} \times Z(G_\Gamma)^{(4)}_{M5} 
\ee
Given flux operators $\Phi_{M2,a}$ with $a\in \Tor H^4(\cM_{10})$ and
$\Psi_{M5,b}$ with $b\in \Tor H^7(\cM_{10})$ (measuring torsional M2 and
M5 charge, respectively) we have that \cite{Freed:2006ya,Freed:2006yc}
\begin{equation}
  \label{eq:11d-commutation-relations}
  \Phi_{M2,a}\Psi_{M5,b} = e^{2\pi i \sL(a,b)}\Psi_{M5,b}\Phi_{M2,a}
\end{equation}
where
\begin{equation}
  \sL \colon \Tor H^4(\cM_{10}) \times \Tor H^7(\cM_{10}) \to \bQ/\bZ
\end{equation}
is the linking pairing in $\cM_{10}$. The
space $\bC^2/\Gamma$ is a cone over $S^3/\Gamma$, and $\Gamma$ acts
freely on the $S^3$, therefore in the case at hand
$\cM_{10}=\cM_7\times (S^3/\Gamma)$. Assuming that $\cM_7$ has no
torsion we can apply the Künneth formula
\begin{equation}
  H^n(\cM_7\times (S^3/\Gamma)) = \sum_{i+j=n} H^i(\cM_7)\otimes H^j(S^3/\Gamma)\, .
\end{equation}
Since
\begin{equation}
  H^\bullet(S^3/\Gamma) = \{\bZ, 0, \Gamma^{\text{ab}}, \bZ\}\, ,
\end{equation}
this implies that
\begin{equation}
  \Tor H^4(\cM_{10}) = H^2(\cM_7)\otimes \Gamma^{\text{ab}} = H^2(\cM_7;\Gamma^{\text{ab}})
\end{equation}
and
\begin{equation}
  \Tor H^7(\cM_{10}) = H^5(\cM_7)\otimes \Gamma^{\text{ab}} = H^5(\cM_7; \Gamma^{\text{ab}})\, .
\end{equation}
Writing, accordingly, $a=\alpha\otimes \ell_a$ and
$b=\beta\otimes \ell_b$, with $\alpha\in H^2(\cM_7)$,
$\beta\in H^5(\cM_7)$ and $\ell_i\in H^2(S^3/\Gamma)=\Gamma^{\text{ab}}$,
we have
\begin{equation}
  \label{eq:7d-commutation-relations}
  \sL(a,b) = (\alpha\cdot \beta) \sL_\Gamma(\ell_1,\ell_2)
\end{equation}
with $\sL_\Gamma$ the linking form in $S^3/\Gamma$. The general form
for $\sL_\Gamma$ is given in \cite{Hikami:2009ze} -- see table
\ref{ADE-linking-forms}. For instance consider the case
$\Gamma=\bZ_N$, corresponding to the $\fsu(N)$ theories in seven
dimensions. We have $\Gamma^{\text{ab}}=[\bZ_N,\bZ_N]=\bZ_N$, with a
linking form
\begin{equation}
  \sL_\Gamma(1,1) = \frac{1}{N} \mod 1\, .
\end{equation}

\begin{table}
$$  
  \arraycolsep=4.4pt\def\arraystretch{1.2}
  \begin{array}{cccc}
    \Gamma & G_\Gamma & \Gamma^{\text{ab}} &  \sL_\Gamma \\
    \hline
    \bZ_N & SU(N) & \bZ_N & \frac{1}{N} \\
    \mathrm{Dic}_{(4N-2)} & \Spin(8N) & \bZ_2\oplus \bZ_2 & \left(\begin{matrix}0 & 1/2 \\ 1/2 & 0\end{matrix}\right) \\
    \mathrm{Dic}_{(4N-1)} & \Spin(8N+2) & \bZ_4 & \frac{3}{4} \\
    \mathrm{Dic}_{(4N)} & \Spin(8N+4) & \bZ_2\oplus \bZ_2 & \left(\begin{matrix}1/2 & 0 \\ 0 & 1/2\end{matrix}\right)\\
    \mathrm{Dic}_{(4N+1)} & \Spin(8N+6) & \bZ_4 & \frac{1}{4} \\
    2T & E_6 & \bZ_3 & \frac{2}{3} \\
    2O & E_7 & \bZ_2 & \frac{1}{2} \\
    2I & E_8 & 0 & 0
  \end{array} 
$$
\caption{Linking pairings for $G_\Gamma$}\label{ADE-linking-forms}
\end{table}

\noindent This implies that the flux operators in this theory will not commute, indicating the defect group of this theory suffers from a mixed 't Hooft anomaly. 
In particular, this entails that some care is needed when choosing boundary conditions. Choose a
basis of the Hilbert space that diagonalizes $\Phi_a$, for
instance.\footnote{That is, choose a basis for the Hilbert space which
  is the unique representation, by the Stone-von Neumanm theorem, of
  the Heisenberg group generated by the flux operators. We refer the
  reader to \cite{Mumford:1338263} for further background material
  on this topic, and to \cite{Tachikawa:2013hya} for a clear
  illustration is a related context.}  Namely:
\begin{equation}
  \Phi_a\ket{b} = e^{2\pi i \sL(a,b)} \ket{b}
\end{equation}
with $b\in \Tor H^7(\cM_{10})$. On the other hand, the states in this
basis do not diagonalize $\Psi_b$
\begin{equation}
  \Psi_{b'}\ket{b} = \ket{b+b'}\, .
\end{equation}
There is analogously a basis of states $\ket{a}$, with
$a\in \Tor H^4(\cM_{10})$ that diagonalizes the $\Psi_b$
operators. The two choices for the basis are related by a discrete
Fourier transform:
\begin{equation}
  \ket{a} = \sum_{b} e^{2\pi i \sL(a,b)}\ket{b}\, .
\end{equation}
For instance, consider choosing a state
$\ket{0,L_{M5}}$ such that $\Psi_b\ket{0,L_{M5}}=\ket{0,L_{M5}}$ for
all $b\in H^7(\cM_{10})$. This corresponds to setting all M5-brane
fluxes to 0, so that the M2 branes are genuine operators. In the seven
dimensional theory we can interpret this choice as being in the
$SU(N)$ theory, with no background fluxes for the 1-form symmetry of
this theory turned on, where the line operators coming from the
wrapped M2 branes are genuine.

If, on the other hand, we choose our boundary conditions to be given
by a state $\ket{0,L_{M2}}$ such that
$\Phi_a\ket{0,L_{M2}}=\ket{0,L_{M2}}$ for all $a\in H^4(\cM_{10})$,
then our background will be in a superposition of all possible
background fluxes for the M5-brane charge, since due to the properties
of the Heisenberg algebra a change of basis from the electric to the
magnetic basis is a discrete Fourier transform:
\begin{equation}
  \ket{0,L_{M2}} = \sum_{a\in H^4(\cM_{10})} \ket{a,L_{M5}}\, .
\end{equation}
In terms of the seven dimensional theory, this implies being on a
superposition of all possible values for the Stiefel-Whitney classes,
or in other words choosing the $SU(N)/\bZ_N$ global form for the
theory, having gauged the one-form symmetry of the $SU(N)$ theory. Notice that, as a consequence of this gauging, the resulting theory has a magnetic $\mathbb Z_N^{(4)}$ higher symmetry

\section{Higher form symmetries and non-commutative fluxes for 5d SCFTs}
\label{sec:5d-general}

The story is fairly similar for five dimensional theories engineered
from M-theory on singular Calabi-Yau threefolds, but the possibilities
in geometry and field theory are much richer. For concreteness, we
will focus on geometries of the kind $\cM_5\times \mathcal V_6$, where $\cM_5$ is
a closed Spin manifold without torsion, and $\mathcal V_6$ a non-compact
Calabi-Yau threefold, given by the cone over some Sasaki-Einstein
manifold $Y_5$. In order to make further progress below we will assume
some additional conditions on $\mathcal V_6$, namely that $H_{2n+1}(\mathcal V_6)=0$ for all
$n$, and that $\Tor H_{2n}(\mathcal V_6)=0$ for all $n$. Toric Calabi-Yau
threefold varieties are an important class of examples that satisfy
these requirements, and we will focus mostly on these below (but the
ideas generalize straightforwardly to other cases, such as cones over
higher del Pezzo surfaces).

\subsection{Field theory analysis}

Let us first describe the expectations from field theory, in analogy
with the seven-dimensional discussion above. In the five dimensional
theory we will have line and surface operators, which we will call
Wilson lines and 't Hooft surfaces, following the standard terminology
in the cases with a Lagrangian description. The charge of these
objects is measured on three-dimensional and two-dimensional surfaces
linking the respective objects in $\cM_5$. Equivalently, depending on
the global structure that we choose for the theory, we have 1-form
symmetries with background fluxes valued on $H^2(\cM_5;\sZ)$, 2-form
symmetries with background fluxes valued on $H^3(\cM_5;\sZ)$, or
combinations of both. Here
$\sZ$ is a group that in the cases with a Lagrangian is given by the
subgroup of the universal cover of the gauge group that leaves all
point operators invariant, as in \cite{Aharony:2013hda}. For instance,
if we have a 5d Yang-Mills theory with algebra $\fsu(N)$ and matter in
the adjoint, we have that $\sZ=\bZ_N$.

As in the seven dimensional case not all of these symmetries are
simultaneously present in any given theory, and one cannot
independently introduce background fluxes for all of them. Rather, we
need to choose a maximal mutually local set of extended operators, and
introduce fluxes only for those. In the Lagrangian context, this
choice is a choice of the global form for the gauge group. For
instance, if the algebra is $\fsu(N)$, a possible choice of global
form is given by $SU(N)/\bZ_N$, where we sum over all background
fluxes in $H^2(\cM_5;\bZ_N)$ --- that is, we sum over Stiefel-Whitney
classes. Alternatively we could sum over bundles in
$H^3(\cM_5;\bZ_2)$. (As remarked above, the sum is not visible in the
usual Lagrangian presentation.)

An interesting feature of five dimensional theories is that instanton
configurations behave very much like particles. In the presence of a
Chern-Simons coupling (for group theory reasons, this coupling is only
available for $\fsu(N)$ with $N>2$)
\begin{equation}
  \label{eq:CS-coupling}
  S_{\text{CS}} = k \int \Omega_A
\end{equation}
where $\Omega_A$ is the Chern-Simons form, these particles can
potentially acquire a charge under the center of the gauge group. If
this happens, then the higher form symmetry of the $SU(N)$ theory can
be (partially) broken, in the same way that ordinary matter in generic
representations breaks the symmetry. Our task below will be to compute
the charge of these particles under the center of the $SU(N)$ gauge
group, but the form of the coupling~\eqref{eq:CS-coupling} suggests
that the right answer will be that in the presence of such a coupling
instanton particles acquire a charge $k$ under the $\bZ_N$ center of
the $SU(N)$ theory. One heuristic way to argue for this is that an
instanton background becomes, in the point-like limit
$\Tr(F^2)=\delta^4(\vec{x})$, with $\vec{x}$ the directions transverse
to the instanton, so the $\Tr(A\wedge F^2)$ term in $\Omega_A$ becomes
an integral of $A$ over the worldline of the instanton particle, so
the Chern-Simons level $k$ can be identified with the charge of the
particle. We will give below a more careful argument that shows that
this is indeed the right result in the theories that we study. This
implies that the center of a $SU(N)$ theory at level $k$ is broken
down to $\bZ_{\gcd(N,k)}$ due to the charge of the instanton
particles.\footnote{Denote by $p$ the order of the surviving group. By
  Lagrange's theorem $p$ divides $N$, so the subgroup is generated by
  $N/p$. For this element to leave a particle of charge $k$ invariant
  we need that $kN/p\equiv 0$ mod $N$, or equivalently that $p$
  divides $k$.\label{fn:gcd(p,q)}}

The case of $SU(2)$ is somewhat special: in this case $\Omega_A$
vanishes identically, but we can introduce a discrete $\theta$-term,
given by 0 or 1 multiplied by the mod-2 index (or equivalently, the
$\eta$ invariant) of a fundamental fermion coupled to the $SU(2)$
bundle that we are introducing. This coupling has the same effect as
above: whenever the coefficient of the coupling is nonzero we find
that the center of $SU(2)$ is broken by the charge of instanton
particles. (We will again give a detailed argument for this fact
below.) This observation generalizes to the $\Sp(2N)$ theories: in
these cases there is a nontrivial bordism group of Spin manifolds
decorated with principal $\Sp$ bundles,
$\Omega_5^{\Spin}(B\Sp(2N))=\bZ_2$ (see for instance
\cite{Garcia-Etxebarria:2018ajm} for a simple derivation). The
non-trivial element in this bordism group can be detected by computing
the $\eta$ invariant of a fermion in the fundamental representation,
so one can couple a $\Sp(2N)$ theory to the SPT defined by this
bordism invariant to introduce a discrete $\theta$ angle. Below we
will only check explicitly that this coupling to a TQFT induces a
charge under the $\bZ_2$ center for instanton particles for $SU(2)$,
but this is actually enough to show that this must also happen for
$\Sp(2N)$ with $N>1$, since in 5 dimensions every $\Sp(2N)$ bundle can
be reduced to a $\Sp(2)$ bundle,\footnote{The obstruction for reducing
  $\Sp(2N+2)$ bundles to $\Sp(2N)$ bundles in $d$ dimensions lives in
  $\pi_i(\Sp(2N+2)/\Sp(2N))$ for $i<d$ (see \cite{Witten:1985bt} for a
  review of the relevant facts for physicists). But
  $\Sp(2N+2)/\Sp(2N)\approx S^{4N+3}$, so the obstructions
  to reducing to $\Sp(2)$ vanish for $d<8$.} and the centers of
$\Sp(2N)$ and any $\Sp(2)$ subgroup can be identified canonically.

\subsection{M-theory and higher form symmetries}

\label{sec:5d-M-theory}

We now want to understand the previous gauge theory discussion in
terms of the M-theory engineering of the relevant 5d theories. The
line and surface operators of the five dimensional operator will come
from M2 and M5 branes wrapping suitable non-compact cycles in the
internal toric Calabi-Yau threefold. As in the seven dimensional case,
we classify which of these operators can be simultaneously taken to be
genuine by looking to a maximal choice of commuting fluxes on the
boundary $\partial\cM_{11}$ (which, recall, has topology
$\cM_5\times Y_5$ in our case, with $Y_5$ a Sasaki-Einstein
manifold). The non-trivial part of the flux commutation
relations will come from the pairing
\begin{equation}
  \sL\colon \Tor H^4(\partial\cM_{11})\times \Tor H^7(\partial\cM_{11}) \to \bQ/\bZ\, .
\end{equation}
We will show momentarily that $Y_5$ only has torsion in
$H_1(Y_5)=H^4(Y_5)$ (or equivalently, by the universal coefficient
theorem \cite{Hatcher}, in $H_3(Y_5)=H^2(Y_5)$). Together with the
fact that $\partial\cM_{11}=\cM_5\times Y_5$, with $\cM_5$ torsion-free, this
implies that we can use the Künneth formulas
\begin{subequations}
  \begin{align}
    \Tor H^4(\partial\cM_{11}) & = \biggl(H^2(\cM_5)\otimes \Tor H^2(Y_5)\biggr) \oplus \biggl(H^0(\cM_5)\otimes \Tor H^4(Y_5)\biggr) \, ,\\
    \Tor H^7(\partial\cM_{11}) & = \biggl(H^3(\cM_5)\otimes \Tor H^4(Y_5)\biggr) \oplus \biggl(H^5(\cM_5)\otimes \Tor H^2(Y_5)\biggr)\, .
  \end{align}
\end{subequations}
Poincaré duality, together with the universal coefficient theorem,
implies that $\Tor H^2(Y_5)=\Tor H^4(Y_5)$. For conciseness, let us
define
\begin{equation}
  \sZ\df \Tor H_3(Y_5) = \Tor H^2(Y_5)=\Tor H^4(Y_5) = \Tor H_1(Y_5)\, .
\end{equation}
(In Lagrangian theories $\sZ$ will be the center of the simply
connected group with the given algebra.)  The defect group for these
geometries is \be \mathbb D = \Big(\sZ^{(1)}_{M2} \oplus
\sZ^{(2)}_{M5}\Big) \oplus \Big(\sZ^{(0)}_{M2} \oplus
\sZ^{(4)}_{M5}\Big) \ee Under the assumption that $\cM_5$ is
torsion-free, the universal coefficient theorem implies that
$H^i(\cM_5)\otimes \sZ=H^i(\cM_5;\sZ)$, so the first terms on the
right hand side are the ones we had anticipated from our field theory
analysis above. We see that in the M-theory language these cohomology
groups parametrize the flux operators measuring the fluxes created by
M2 branes wrapping non-compact two-cycles in $V$, and M5 branes
wrapping non-compact four-cycles in $V$, respectively, as one would
have expected. The last two terms correspond to $(-1)$-form and 4-form
symmetries, which are somewhat more exotic from the field theory point
of view, and we will ignore them in our analysis. (See
\cite{Sharpe:2019ddn,Tanizaki:2019rbk,Gu:2020ivl} for recent work
exploring such symmetries from the field theory point of view.)

\medskip

For our purposes it will be convenient to work with a smoothed out and
compactified version of the geometry. Denote by $\tilde {\mathcal V_6}$ some smooth
crepant resolution of $\mathcal V_6$ (it does not matter which one), and
introduce $X_6\df B_6 \cap \tilde {\mathcal V_6}$, where $B_6$ is a sufficiently
large ball containing the exceptional set of $\tilde {\mathcal V_6}$. Branes
extending from the origin to infinity in $\mathcal V_6$ will map to branes
extending to the boundary of $X_6$. We have that
$H_i(\tilde {\mathcal V_6})=H_i(X_6)$, and since (topologically) $Y_5=\partial X_6$
we have a long exact sequence of the form
\begin{equation}
  \label{eq:relative-homology-LES}
  \ldots \to H_n(Y_5) \to H_n(X_6) \to H_n(X_6,Y_5) \to H_{n-1}(Y_5) \to \ldots
\end{equation}
The physical interpretation of this long exact sequence in our
physical context is as follows: branes wrapped on $H_n(X_6)$ give rise
to dynamical objects in the theory. Branes going to infinity will be
associated to elements of the relative homology groups $H_n(X_6,Y_5)$:
these are defined to be those cycles in $X_6$ that are closed, modulo
cycles living on $Y_5$. These branes wrapping non-compact cycles give rise to defects in the theory, whose lattice of charges is given precisely by this group. However, physically, we are only interested in the charges that
survive 't Hooft screening by dynamical operators, mathematically this is
encoded in the fact that we only care about the quotient
$H_n(X_6,Y_5)/H_n(X_6)$, or equivalently that we only need to know
about the homology class of the intersection of the non-compact cycle
with the boundary, namely $H_{n-1}(Y_5)$. (The long exact sequence
does not necessarily truncate on $H_{n-1}(Y_5)$ in general, so this
statement would need correction in those cases in which it doesn't,
but the truncation does take place in all cases of interest to us.)

\medskip

Next, we use Lefschetz duality \cite{Hatcher} to rewrite
\begin{equation}
    H_k(X_6,Y_5) = H_c^{n-k}(X_6) \, .
\end{equation}
We will assume that $X_6$ is torsion free (which is true, in
particular, for toric varieties \cite{BF,Cox,Dan}), so by the
universal coefficient theorem for cohomology we get
\begin{equation}
    H_{k}(X_6,Y_5)=H^{n-k}_c(X_6)=\Hom (H_{n-k}(X_6), \bZ)\, .
\end{equation}
If we now assume that $H_{2n+1}(X_6)=0$ (which is again true in the
special case of toric varieties \cite{BF,Cox,Dan}, but also more
generally), the long exact sequence~\eqref{eq:relative-homology-LES}
reduces to
\begin{subequations}
  \label{eq:split-relative-homology-LES}
  \begin{align}
    0 \to \Hom (H_0(X_6), \mathbb{Z}) \to H_5(Y_5)\to 0\, ,\\
    0\to H_4(Y_5)\to H_4(X_6) \xrightarrow {Q_4} \Hom (H_2(X_6), \mathbb{Z}) \xrightarrow{\partial_4} H_3(Y_5) \to 0\, ,\\
    0 \to H_2(Y_5) \to H_2(X_6) \xrightarrow{Q_2} \Hom (H_4(X_6), \mathbb{Z}) \xrightarrow{\partial_2} H_1(Y_5) \to 0\, ,\\
    0 \to H_0(Y_5) \to H_0(X_6) \to 0\, ,
  \end{align}
\end{subequations}
where, the homomorphisms $Q_k\colon H_k(X_6)\to \Hom(H_{6-k}(X_6),\bZ)$ are
given by partial evaluation of the intersection forms
\begin{equation}
  q_k\colon H_k(X_6)\times H_{6-k}(X_6)\to \mathbb{Z}
\end{equation}
with $k=2, 4$. That is, $Q_k(x)(y)=q_k(x,y)$. Note that
$Q_4=Q_2^t$. It follows from these exact sequences that
$H_0(Y_5)=H_5(Y_5)=H_0(X)=\bZ$, and that
\begin{equation}\label{eq:homy}
  \begin{split}
    H_4(Y_5)=\ker(Q_4),\quad & H_3(Y_5)= \coker(Q_4),\\
    H_2(Y_5)=\ker(Q_2),\quad &
    H_1(Y_5)=\coker(Q_2),
  \end{split}
\end{equation}
so finally
\begin{equation}
  \sZ = \Tor \coker(Q_4) = \Tor \coker(Q_2)\, .
\end{equation}

Having understood the space of charge operators for the five
dimensional theory, we still need to find their commutation
relations. This follows straightforwardly
from~\eqref{eq:11d-commutation-relations}, in a way very analogous
to~\eqref{eq:7d-commutation-relations}. Writing
$a=\alpha\otimes \Sigma_a$ and $b=\beta\otimes D_b$, with
$\alpha\in H^2(\cM_5)$, $\beta\in H^3(\cM_5)$,
$\Sigma_a\in \Tor H^2(Y_5)$ and $D_b\in \Tor H^4(Y_5)$, we have
\begin{equation}
  \label{eq:5d-commutation-relation}
  \sL(a,b) = (\alpha\cdot \beta) \sL_{Y_5}(\Sigma_a,D_b)
\end{equation}
with $\sL_{Y_5}$ the linking form in $Y_5$, a perfect pairing
\begin{equation}
  \sL_{Y_5}\colon \Tor H^k(Y_5)\times \Tor H^{5-k+1}(Y_5)\to \mathbb{Q}/\mathbb{Z}\, .
\end{equation}
We can derive $L_{Y_5}$ from knowledge of the intersection matrix
$Q_4=Q_2^t$ as follows.
Let $\sigma\in\Tor H_{3}(Y_5)$ and $\bar{\sigma}\in\Tor H_{1}(Y_5)$,
and choose $\mu\in\Hom(H_2(Y_5),\bZ)$ and
$\bar{\mu}\in\Hom(H_4(Y_5),\bZ)$ such that, $\partial_4 \mu=\sigma$
and $\partial_2 \bar{\mu}=\bar{\sigma}$. Then, for non-trivial
$\Tor H_{3}(Y_5)$ and $\Tor H_{1}(Y_5)$, there are non-zero integers
$n$ and $m$ such that $\partial (n\mu)=n\sigma=0$ and
$\bar{\partial} (m\bar{\mu})=m\bar{\sigma}=0$. Thus, we may pick
$\nu\in H_4(X_6)$ and $\bar{\nu}\in H_2(X_6)$ such that, $Q_4\nu=n\mu$
and $Q_2\bar{\nu}=m\bar{\mu}$. The linking pairing is
then\footnote{Note that we use $q$ and $q^t$ interchangeably. It
  should be clear from the context which one we mean.} \cite{FS}
\begin{equation}\label{link1}
  \sL_{Y_5}(\sigma,\bar{\sigma})\equiv\frac{1}{nm}q(\nu,\bar{\nu})\quad \mod 1\, .
\end{equation}
This may be equivalently written as 
\begin{equation}\label{link2}
  \sL_{Y_5}(\sigma,\bar{\sigma})\equiv q^{-1}(\mu,\bar{\mu})\mod 1\, ,
\end{equation}
where
$q^{-1}\colon \Hom (H_2(X_6), \bZ)\times \Hom (H_4(X_6), \bZ)\to
\bQ$. More explicitly, this means that, if $\alpha^{'*}_i$ is a
generator of $\Hom (H_2(X_6), \bZ)$ and $\beta^{'*}_j$ is a generator
of $\Hom (H_4(X_6), \bZ)$ such that $\partial\alpha^{'*}_i$ is the
generator of $\Tor H_{3}(Y_5)$ and $\bar{\partial}\beta^{'*}_j$ is the
generator of $\Tor H_{1}(Y_5)$ then, the linking number is just the
$(i\times j)$th element of $q^{-1}$:
\begin{equation}\label{link3}
  \sL_{Y_5}(\partial\alpha^{'*}_i,\bar{\partial}{\beta^{'*}_j})=
  q^{-1}(\alpha^{'*}_i,{\beta^{'*}_j})=
  q^{-1}_{ij}
  \mod 1\, .
\end{equation}
The appendices contain various worked out examples of the application
of this relation, which encodes the discrete mixed 't Hooft anomaly coefficients for defect groups associated to the higher form symmetries of 5d SCFTs.

\subsection{The case of toric Calabi-Yau varieties}

\label{sec:toric-general}

An important special case of the previous discussion is that when
$\mathcal V_6$ is a toric Calabi-Yau variety. (We refer the reader to
\cite{CLS,Ref,Denef:2008wq,Hori:2003ic} for systematic reviews of
toric geometry.)

The crepant resolution $\tilde {\mathcal V_6}$ is obtained by choosing a
triangulation for the toric diagram. As mentioned above, the odd
dimensional homology groups of a toric variety vanish, and the even
homology groups for $X_6$ can be easily obtained by looking at the
toric diagram for $\mathcal V_6$. Let $I$ be the number of points in the interior
of the diagram and $B$ be the number of points on the edges of the
diagram. Then the number of $4$-cycles is $I$, and since $\mathcal V_6$ is
connected the number of $0$-cycles is $1$.  The Euler characteristic
of the Calabi-Yau equals to the number of $2$-dimensional faces of the
resolved toric diagram \cite{He17}, which is twice the area $A$ of the
toric diagram. The Euler characteristic also equals the number of
even-dimensional cycles minus the number of odd-dimensional cycles. We
know by Pick's theorem that $2A= 2I +B -2$, so the number of
$2$-cycles is $I+B-3$, so we have
\begin{equation}
  \label{eq:toric-homology}
  H_0(X)=\mathbb{Z},\; H_2(X)=\mathbb{Z}^{(I+B-3)},\; H_4(X)=\mathbb{Z}^I,\;
  H_6(X)=0\, .
\end{equation}

One can also compute $Q_4$ (or equivalently $Q_2$) in toric varieties
quite conveniently purely in terms of the toric data. What one needs
to construct is the Mori cone of effective curves in the toric
variety, and find their intersections with the compact divisors, which
are manifest in the toric description as points in the interior of the
toric diagram. Well developed algorithms for doing this exist,
reviewed for example in \cite{Ref}, and implemented for instance in
\textsc{Sage} \cite{sage}. As an example, consider the Calabi-Yau cone
over $\bF_0=\bP^1\times\bP^1$. This geometry can be alternatively
described as the (real) Calabi-Yau cone over $Y^{2,0}$. Its toric
diagram has external vertices $p_i=\{(1,0),(-1,0),(0,1),(0,-1)\}$, and
an internal vertex at $t=(0,0)$. Its Mori cone is generated by two
curves $\cC_1$, $\cC_2$ with intersection matrix with the toric
divisors given by
\begin{equation}
  \begin{array}{c|ccccc}
    & V(p_1) & V(p_2) & V(p_3) & V(p_4) & V(t)\\
    \hline
    \cC_1 & 1 & 0 & 1 & 0 & -2\\
    \cC_1 & 0 & 1 & 0 & 1 & -2
  \end{array}
\end{equation}
We have included all toric divisors here, but the $V(p_i)$ divisors
are non-compact. The divisor $V(t)$ is compact, on the other hand, so
we find that
\begin{equation}
  Q_2 = \begin{pmatrix}
    -2 & -2
  \end{pmatrix}
\end{equation}
and we predict that
\begin{equation}
  \sZ = \Tor \coker(Q_2) = \bZ_2\, .
\end{equation}
We will see below that this result agrees with the expectation from
field theory: the defect group in this case is
\be
\mathbb D = \bZ_2^{(1)} \oplus \bZ_2^{(2)}.
\ee

\medskip

All the results below can be derived using these methods, but in
practice it is much more efficient to use instead a method introduced
(to our knowledge) in \cite{Garcia-Etxebarria:2016bpb}, which avoids
the need to introduce a triangulation or computing the Mori
cone. Consider a toric Calabi-Yau cone with an isolated singularity,
and $v$ external vertices. (In \S\ref{sec:non-isolated-prescription}
we will derive a modified form of the method in
\cite{Garcia-Etxebarria:2016bpb} valid for non-isolated
singularities.) In terms of the toric diagram, this means that there
are no lattice points along the edges of the toric diagram. As argued
in \cite{Garcia-Etxebarria:2016bpb}, one has that
\begin{equation}
  H_{i}(Y_5) = H_i(B_3^L) \qquad \text{for } i\leq 2\, ,
\end{equation}
where $B_3^L$ is a chain of lens spaces $L_{n_1},\ldots,L_{n_v}$,
joined at their torsion cycle, constructed as follows. For each
external vertex $p_i$, $i\in\{1,\ldots,v\}$, construct the triangle
$T_i$ defined by the vertex and the two vertices adjacent to it, that
is, the convex hull of $\{p_{i-1},p_i,p_{i+1}\}$ (with $p_0\df p_v$
and $p_{v+1}\df p_1$). Then $n_i=2\mathrm{Area}(T_i)$. Additionally,
one can show that \cite{Garcia-Etxebarria:2016bpb}
\begin{equation}
  H_1(B_3^L) = \bZ_{\gcd(n_1,\ldots,n_v)}\, ,
\end{equation}
so we find that in the toric case
\begin{equation}
  \label{eq:toric-torsion}
  \sZ = \bZ_{\gcd(n_1,\ldots,n_v)}
\end{equation}
Coming back to our $\cC_\bR(Y^{2,0})$ example, we have four triangles,
all of unit area. So
\begin{equation}
  \sZ=\bZ_{\gcd(2,2,2,2)}=\bZ_2\, .
\end{equation}

\begin{figure}
\begin{center}
  \begin{minipage}{2cm}
    $\begin{gathered}\xymatrix@=0.5em{&\bullet_{v-1}\ar@{-}[r]\ar@{-}[dl]&\cdots\\ \bullet_v\ar@{-}[d]\\ \bullet_1\ar@{-}[dr]&&&\vdots\\&\bullet_2\ar@{-}[r]&\bullet_3\ar@{-}[ur]}\end{gathered}$
    \vspace{3cm}
  \end{minipage}
  \hspace{3cm} \includegraphics[scale=0.6]{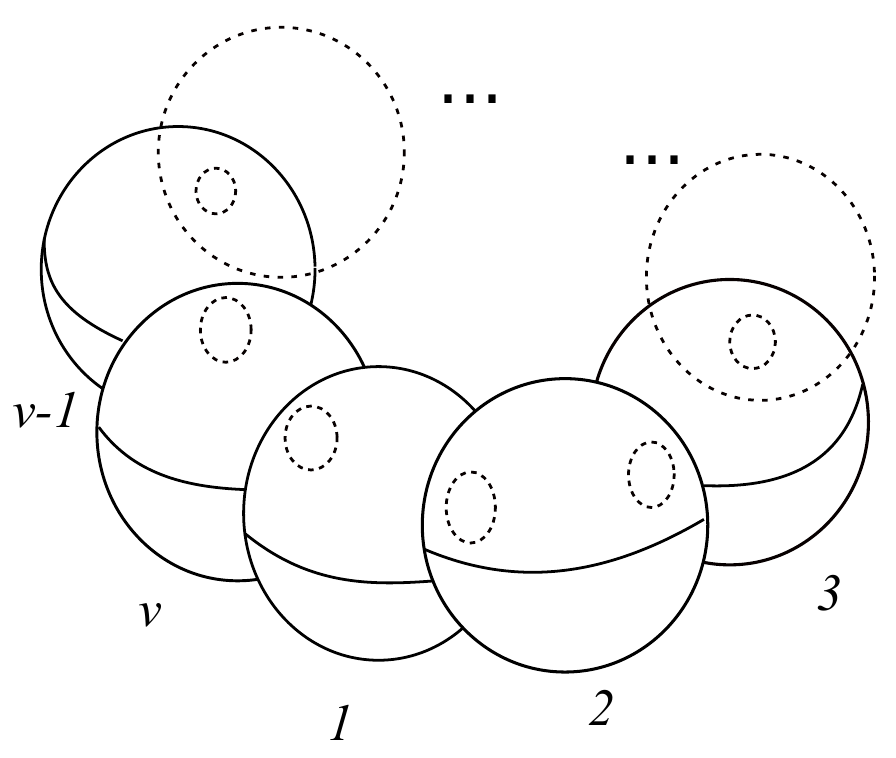}
  \vspace{-3cm}
\end{center}
\caption{Schematic topology of $B_3^L$ \cite{Garcia-Etxebarria:2016bpb}}
\end{figure}

\section{$\fsu(p)_k$ theory}

\label{sec:su(p)_q}

We will now apply the previous results to a simple set of cases:
$\fsu(p)$ theories at Chern-Simons level $k$.

\subsection{Toric realization and geometric computation}

It is well-known that $\fsu(p)_k$ theories can be obtained exploiting
canonical CY singularities that are cones over Sasaki-Einstein
manifolds of $Y^{p,q}$ type (see, for instance, \cite{Closset:2018bjz}
for a recent account and \cite{Gauntlett:2004yd} for the original
analysis of these geometries).  Let us introduce for
convenience $q\df -k$, and we will assume $0\leq q < p$. We show the
resulting toric diagram in figure~\ref{fig:Y^{p,q}-toric}.

\begin{figure}
  \centering
  $$\xymatrix@=0.7em{
  &\overset{(0,p)}{\bullet}&\\
  &\bullet&\\
  &\vdots&\underset{(1,l)}{\bullet}\ar@{-}[uul]\\
  &\bullet&\\
  \underset{(-1,0)}{\bullet}\ar@{-}[r]\ar@{-}[uuuur]&\underset{(0,0)}{\bullet}\ar@{-}[uur]&\\
  }$$
  \caption{Toric diagram for $Y^{p,q}$. We have defined $l \equiv p-q$.}
  \label{fig:Y^{p,q}-toric}
\end{figure}
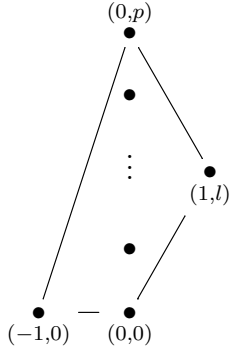

From our general discussion above, we need to compute
$\Tor H^2(Y^{p,q}) = \Tor H^4(Y^{p,q})$, together with the linking
pairing, in order to determine the Heisenberg group encoding the
higher symmetries of the theory. Whenever $p$ and $q$ are relatively
prime, we have that \cite{Gauntlett:2004yd} $Y^{p,q}$ is topologically
$S^2\times S^3$, so there is no torsion. So in these cases there is no
choice of global structure for the field theory. More interesting is
the case where $\gcd(p,q)\neq 1$. We can compute the relevant torsion
groups following the general prescription in \S\ref{sec:toric-general}
as follows. Choose an ordering of the external points of the toric
diagram in figure~\ref{fig:Y^{p,q}-toric} such that adjacent points
are consecutive. For instance, choose
\begin{equation}
  \{p_1,p_2,p_3,p_4\} = \{(-1,0), (0,0), (1,l), (p,0)\}\, .
\end{equation}
Define now the triangles $T_i$, $i\in\{1,\ldots,4\}$, as the convex
hull of $\{p_{i-1},p_i,p_{i+1}\}$ (with $p_{0}\df p_4$ and
$p_{5}\df p_1$). We show the triangle $T_2$ in
figure~\ref{fig:Y^{p,q}-torsion} as an example.
\begin{figure}
  \centering
  $$\xymatrix@=0.7em{
  &\overset{(0,p)}{\bullet}&\\
  &\bullet&\\
  &\vdots&\underset{(1,l)}{\bullet}\ar@{-}[uul]\\
  &\bullet&\\
  \underset{(-1,0)}{\bullet}\ar@{-}[r]\ar@{-}[uuuur]\ar@{..}[uurr]&\underset{(0,0)}{\bullet}\ar@{-}[uur]&\\
  }$$
  \caption{One of the triangles defined in the text.}
  \label{fig:Y^{p,q}-torsion}
\end{figure}
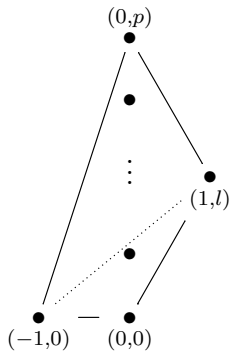
We have that
\begin{equation}
  \Tor H^2(Y^{p,q}) = \Tor H^4(Y^{p,q}) = \bZ_{\gcd(n_1,n_2,n_3,n_4)}
\end{equation}
where $n_i$ is defined as twice the area of $T_i$. It is elementary to
show that $n_i=\{p,l,p,2p-l\}=\{p,p-q,p,p+q\}$, which implies that
\begin{equation}
  \sZ=\Tor H^2(Y^{p,q}) = \Tor H^4(Y^{p,q}) = \bZ_{\gcd(p,q)}\, .
\end{equation}
We show in appendix~\ref{app:L_{Y^{p,q}}} that the linking pairing
$\sL_{Y^{p,q}}\colon \Tor H^2(Y^{p,q})\times \Tor H^4(Y^{p,q}) \to \mathbb Q / \mathbb Z$ is
\begin{equation}
  \sL_{Y^{p,q}}(1,1) = -\frac{1}{\gcd(p,q)} \mod 1\, .
\end{equation}

In the case that the Chern-Simons level $k$ vanishes this leads to
$\sZ=\bZ_p$, which is the expected result for pure $\cN=1$ $\fsu(p)_0$
theory in five dimensions. This theory admits a number of global
variants, for instance $SU(p)_0$ or $PSU(p)_0\df SU(p)_0/\bZ_p$. The
classification of all such global forms proceeds just as in the case
of $\fsu(p)$ theories in four dimensions \cite{Aharony:2013hda}, so we
will not delve on it further. The case with $k\neq 0$ is more subtle,
and we turn to it now.

\subsection{Non-vanishing Chern-Simons levels and the charge of instanton particles}

We would now like to reproduce the M-theory results above from a field
theoretical perspective, particularly in the case $q>0$. We will see
that the instanton particles are charged under the center $\bZ_p$ of
the gauge group, so they will break this group to the subgroup under
which the instanton particles are uncharged.

In order to determine the charge of the instanton particles, note that
the instanton particles will arise in M-theory from M2 branes wrapping
holomorphic curves in the geometry. On the other hand, the center of
the gauge group lies in the orbit of the Cartan generators, which can
be understood geometrically in terms of divisors in the M-theory
picture. So, in this context, the charge of the instanton particles
will be encoded in an intersection of the effective curve that the
instanton is wrapping with some suitably chosen combination of
divisors.

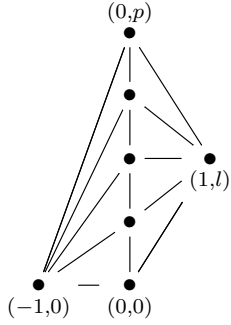
\begin{figure}
  \centering
  $$\xymatrix@=0.7em{
  &\overset{(0,p)}{\bullet}&\\
  &\bullet\ar@{-}[u]&\\
  &\bullet\ar@{-}[u]&\underset{(1,l)}{\bullet}\ar@{-}[uul]\ar@{-}[ul]\ar@{-}[dl]\ar@{-}[l]\ar@{-}[ddl]\\
  &\bullet\ar@{-}[d]\ar@{-}[u]&\\
  \underset{(-1,0)}{\bullet}\ar@{-}[r]\ar@{-}[ur]\ar@{-}[uur]\ar@{-}[uuur]\ar@{-}[uuuur]\ar@{-}[uuuur]&\underset{(0,0)}{\bullet}\ar@{-}[uur]&\\
  }$$
  \caption{Triangulation of $Y^{p,q}$ considered in the text.}
  \label{fig:Y^{p,q}-triangulation}
\end{figure}

It is useful to choose the triangulation of the $\cC(Y^{p,q})$
geometry given in figure~\ref{fig:Y^{p,q}-triangulation}. Denoting the
points at $(0,i)$ by $I_i$, the cones of this triangulation are of the
form $(p_1, I_i, I_{i+1})$ and $(p_3, I_i, I_{i+1})$. The instanton
particles are naturally associated to the curves
$\cC_i\df D_{I_{i}}\cdot D_{I_{i-1}}$, with $D_{I_i}$ the toric
divisor associated to the point $I_i$, and $i\in \{1,\ldots,p\}$. On
the other hand, the Cartan generators are associated to the interior
points: for each compact four-cycle $D$ in the geometry we have a
Poincaré dual harmonic two-form $\PD[D_i]$, and dimensional reduction
of $C_3$ along these harmonic forms gives rise to the gauge bosons in
the Cartan of the five dimensional theory. In the geometry at hand the
compact four-cycles are generated by the toric divisors associated to
the interior points $I_i$, $i\in\{1,\ldots,p-1\}$. The charge of a
curve $\cC$ under the $U(1)$ Cartan associated to the divisor is then
simply
\begin{equation}
  Q_i(\cC) = \int_{\cC} \PD[D_{I_i}] = D_{I_i}\cdot \cC\, .
\end{equation}

We are only interested on a very specific element of the Cartan of
$SU(p)$, the generator of the $\bZ_p$ center of the group. This
generator is of the form
\begin{equation}
  g_Z = \begin{pmatrix}
    \omega_p \\
    & \omega_p \\
    && \ddots \\
    &&& \omega_p
  \end{pmatrix}
\end{equation}
with $\omega_p=\exp(2\pi i/p)$. On the other hand, it is natural to
choose an embedding of the $U(1)$ subgroups associated to the $D_i$
into $SU(p)$ of the form
\begin{equation}
  g_1(\alpha) = \begin{pmatrix}
    e^{i\alpha} \\
    & e^{-i\alpha}\\
    && 1 \\
    &&& 1\\
    &&&& \ddots
  \end{pmatrix} \quad ; \quad
  g_2(\alpha) = \begin{pmatrix}
    1 \\
    & e^{i\alpha}\\
    && e^{-i\alpha} \\
    &&& 1 \\
    &&&& \ddots
  \end{pmatrix} \quad \ldots \quad
  g_{p-1}(\alpha) = \begin{pmatrix}
    1 \\
    & \ddots \\
    && 1\\
    &&& e^{i\alpha} \\
    &&&& e^{-i\alpha}
  \end{pmatrix}\, .
\end{equation}
(One reason that this embedding is natural is that the $D_{I_i}$
geometrically parametrize separation of the Cartan branes in a $U(p)$
stack, and each Cartan brane in the $U(p)$ theory naturally embed as
$\text{diag}(1,\ldots,1,e^{i\alpha},1,\ldots)$.) We can write
\begin{equation}
  g_Z = \prod_{k=1}^{p-1}\left(g_k(2\pi/p)\right)^{k}\, .
\end{equation}
So if we want to measure the charge of a curve $\cC$ under the
generator $g_Z$ of the center of $SU(p)$, we can define a divisor
$Z\df \sum_{k=1}^{p-1} k D_{I_k}$, and the charge will be given by
\begin{equation}
  Q_Z(\cC) \equiv \cC \cdot Z = \sum_{k=1}^{p-1} k\,\cC\cdot D_{I_k} \mod p
\end{equation}
where we have used the fact that the charge under the centre is
defined modulo $p$.

We thus only need to determine the charge of the instanton particles
associated to $\cC_j$ under the Cartan associated to $D_{I_i}$. We denote
this by $Q_i(\cC_j)$. It is straightforward to compute from the toric
data, and it was also obtained using field theory methods in
\cite{Closset:2018bjz}. Either way, one obtains
\begin{equation}
  \begin{split}
    Q_{i-1}(\cC_i) & = (p - q - 2i) \\
    Q_{i}(\cC_i) & = -(p - q - 2(i-1))
  \end{split}
\end{equation}
with all other charges vanishing. The charge under the centre is thus
\begin{equation}
  Q_Z(\cC_i) \equiv (i-1)(p-q-2i) - i(p-q - 2(i-1)) \equiv q\mod p\, .
\end{equation}
So we obtain the result (that one might have guessed from the form of
the Chern-Simons coupling in the first place) that instanton particles
have charge $q$ under the center of $SU(p)$. Recalling that the
subgroup of $\bZ_p$ preserved by a particle with charge $q$ is
precisely $\gcd(p,q)$, we reproduce the result from the geometric
computation above.

There is one last remaining subtlety to take care of: we have just
shown that there are instantonic particles have charge $q$ mod $p$,
but it could in principle be possible that there is some class of
particles with charge different from $q$ or 0 under $\bZ_p$, which
would change the result. It is not difficult to show that this is not
the case, as follows. Every particle will wrap an effective curve in
the geometry, or in other words a curve in the Mori cone of the toric
variety. This Mori cone is finitely generated, we give a set of
generators in appendix~\ref{app:Mori-cone-Y^{p,q}}. It is a simple
exercise, using the ideas above, to show that all the generators of
the Mori cone found there have charge 0 or $q$ under $\bZ_p$, so the
same will hold for any curve in the Mori cone. So we can conclude that
the $\bZ_{\gcd(p,q)}$ 1-form symmetry is not broken any further by M2
brane states.

\subsection{Adding fundamental matter}

\begin{figure}
  \centering
 $$\xymatrix@=0.7em{
  &\overset{(0,p)}{\bullet}&\\
  &\bullet&\\
  &\vdots&\underset{(1,l)}{\bullet}\ar@{-}[uul]\\
  &\bullet&\bullet\ar@{-}[u]\\
  \underset{(-1,0)}{\bullet}\ar@{-}[r]\ar@{-}[uuuur]&\underset{(0,0)}{\bullet}\ar@{-}[ur]\ar@{..}[uur]&\\
  }$$
  \caption{Geometry realizing $\fsu(p)$ with one flavor. We
    show the added triangle, which is of minimal area.}
  \label{fig:su(p)-one-flavor}
\end{figure}
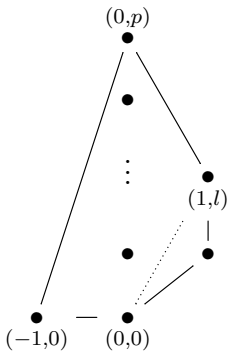

As a final check of our formalism, let us consider the case of the
$SU(p)_q$ theory with a hypermultiplet in the fundamental
representation. This theory can be engineered by considering the
geometry in figure~\ref{fig:su(p)-one-flavor}
\cite{Aharony:1997bh}. It is related by adding a triangle to the
$Y^{p,q}$ toric diagram. Note that the added triangle is of minimal
area, which implies that there is no torsion in the horizon manifold
for this geometry, so we will have that $\sZ$ is trivial, reproducing
the field theory expectation that upon addition of a flavor in the
fundamental representation the higher form symmetry is broken.

\section{Higher form symmetries and 5d duality}\label{sec:5duality}
Often massive deformations of a 5d SCFTs can give rise to gauge theory phases \cite{Seiberg:1996bd}. A given 5d SCFT 
can admit several such phases related to inequivalent effective gauge theories thus resulting in a so called 5d ``duality'' \cite{Aharony:1997ju,Aharony:1997bh} (see also \cite{Bhardwaj:2019ngx} for a huge list of novel such dualities predicted using geometric engineering). Here we use the word duality in quotes to emphasize that this is not a duality in any conventional field theoretical sense, rather the manifestation that the Coulomb phases of certain effective gauge theory description of a relevant deformation of a 5d SCFT happen to overlap \cite{Closset:2018bjz}. Higher form symmetries provide an interesting consistency condition: these deformations are neutral with respect to the higher forms of the SCFTs, and therefore different 5d dual gauge theories must have the same higher form symmetries.

\medskip

In this section we illustrate some examples of this consistency checks
building upon a field theory analysis. As many of these examples have
also nontrivial nonabelian 0-form symmetry groups, the corresponding
geometries typically have corresponding curves of singularities, and
our prescription to compute the center symmetry must be modified to
include non-isolated singularities. Having at our disposal several
gauge theory examples, it is easy to conjecture a modified
prescription that works in this case and reproduces all the field
theoretical results. After motivating the conjecture from physical
considerations, we will prove that it is indeed correct.

\subsection{The beetles}
Consider the 5d duality
\be
SU(2)_\pi \times SU(2)_\pi \,\,\longleftrightarrow\,\, SU(3)_0 \,\, N_f = 2 
\ee
which can be understood from the corresponding geometry as follows. The 5d SCFT from which this 5d duality originates is realized as a $\TX$ where $\sing$ is the toric CY singularity with toric diagram
\be
\xymatrix{
&\bullet\ar@{-}[r]&\bullet\ar@{-}[dr]&\\
\bullet\ar@{-}[ur]\ar@{-}[dr]& \circ & \circ & \bullet\\
&\bullet\ar@{-}[r]&\bullet\ar@{-}[ur]&
}
\ee
the so called `beetle geometry' \cite{Closset:2018bjz}. The $SU(2) \times SU(2)$ phase can be characterized via the vertical reduction/ruling (see \cite{Intriligator:1997pq} for the original analysis and \cite{Closset:2018bjz} for some recent results)
\be
\begin{gathered}
\xymatrix{
&\bullet\ar@{-}[r]\ar@[red]@{-}[d]&\bullet\ar@{-}[dr]\ar@[red]@{-}[d]&\\
\bullet\ar@{-}[ur]\ar@{-}[dr]& \circ\ar@[red]@{-}[d] & \circ\ar@[red]@{-}[d] & \bullet\\
&\bullet\ar@{-}[r]&\bullet\ar@{-}[ur]&
\\
&\overset{2}{\circ}\ar@{-}[r]&\overset{2}{\circ}
}\\
\end{gathered}
\ee
The $SU(3)$ $N_f=2$ phase can be characterized via the horizontal reduction/ruling
\be
\xymatrix{
&\bullet\ar@{-}[r]&\bullet\ar@{-}[dr]&&&\overset{1}{\square}\ar@{-}[d]\\
\bullet\ar@[red]@{-}[r]\ar@{-}[ur]\ar@{-}[dr]& \circ \ar@[red]@{-}[r]& \circ\ar@[red]@{-}[r] & \bullet&&\overset{3}{\circ}\ar@{-}[d]\\
&\bullet\ar@{-}[r]&\bullet\ar@{-}[ur]&&&\overset{1}{\square}
}
\ee
From this latter perspective it is clear that this model cannot have higher form symmetries: the fundamental of $SU(3)$ has unit charge with respect to the center of the gauge group, and therefore the corresponding one form symmetry is broken. However, from the perspective of the $SU(2)\times SU(2)$ phase, it might seem that we still have a one form symmetry: indeed via the isomorphism 
\be
SU(2) \times SU(2)\simeq \text{Spin(4)}
\ee
we can identify the bifundamental $(\mathbf{2},\mathbf{\bar 2})$ as a $\mathbf{4}_v$, which has charge $(1,0)$ with respect to the $\mathbb Z_2 \times \mathbb Z_2$ center of $\text{Spin}(4)$. For this reason one might expect to have a surviving $\mathbb Z_2$ factor. The solution of this conundrum is given by realizing that we have a non-trivial discrete theta term for the two $SU(2)$ factors in this theory. Geometry clearly encodes this fact: recall that the singularity obtained by shrinking a $dP_1$ surface to zero size corresponds to the $SU(2)_{\pi}$ gauge theory, that has a nontrivial discrete theta angle (and no one form symmetry, as we showed in \S\ref{sec:su(p)_q}). Cutting the toric diagram open along the dotted diagonal
\be
\begin{gathered}
\xymatrix{
&\bullet\ar@{-}[r]\ar@[red]@{-}[d]&\bullet\ar@{-}[dr]\ar@[red]@{-}[d]&\\
\bullet\ar@{-}[ur]\ar@{-}[dr]& \circ\ar@[red]@{-}[d] & \circ\ar@[red]@{-}[d] & \bullet\\
&\bullet\ar@{-}[r]\ar@{..}[uur]&\bullet\ar@{-}[ur]&
}
\end{gathered} \qquad \longleftrightarrow \qquad \begin{gathered}
\xymatrix{
&\bullet\ar@{-}[r]\ar@[red]@{-}[d]&\bullet\\
\bullet\ar@{-}[ur]\ar@{-}[dr]& \circ\ar@[red]@{-}[d]\\
&\bullet\ar@{..}[uur]} 
\end{gathered} + \begin{gathered}
\xymatrix{
&\bullet\ar@{-}[dr]\ar@[red]@{-}[d]&\\
& \circ\ar@[red]@{-}[d] & \bullet\\
\bullet\ar@{-}[r]\ar@{..}[uur]&\bullet\ar@{-}[ur]&
}
\end{gathered} 
\ee
we clearly see that we can identify the two gauge theory subsectors with $SU(2)_\pi$ theories that have a trivial $\mathbb Z_2$ one-form symmetry: because of the discrete theta term, the corresponding BPS instantons are charged with respect to the center of the gauge group and break the diagonal $\mathbb Z_2$. This result clearly matches our prescription: the defect group for this geometry is trivial, since this geometry has outer triangles of minimal area.

\bigskip

This argument generalizes to the case of other quiver theories. As an example let's consider the duality
\be
SU(2)_\pi\times \underbrace{SU(2)_0 \times \cdot \times SU(2)_0}_{N-2 \text{ times}} \times SU(2)_\pi \quad \longleftrightarrow \quad SU(N+1) \,\, N_f = 2N-2\,.
\ee
The 5d SCFT from which this 5d duality originates is realized as a $\TX$ where $\sing$ is the toric CY singularity with toric diagram
\be
\begin{gathered}
\xymatrix{
&\bullet\ar@{-}[r]&\bullet\ar@{-}[r]&\cdots\ar@{-}[r]&\bullet\ar@{-}[r]& \bullet\ar@{-}[dr]&\\
\bullet\ar@{-}[ur]\ar@{-}[dr]& \circ & \circ &\cdots& \circ& \circ & \bullet\\
&\bullet\ar@{-}[r]&\bullet\ar@{-}[r]&\cdots\ar@{-}[r]&\bullet\ar@{-}[r]&\bullet\ar@{-}[ur]&
}\\
\quad\underbrace{\phantom{\xymatrix{&&&&&}}}_{N\text{ times}}
\end{gathered}
\ee
From the horizontal reduction it is manifest that we do not have any residual one-form symmetry: we obtain an $SU(N+1)$ theory coupled to matter in the fundamental, which completely breaks the center symmetry, which is manifest from the following horizontal reduction/ruling
\be
\begin{gathered}
\xymatrix{
&\bullet\ar@{-}[r]&\bullet\ar@{-}[r]&\cdots\ar@{-}[r]&\bullet\ar@{-}[r]& \bullet\ar@{-}[dr]&&\overset{N-1}{\square}\ar@{-}[d]\\
\bullet\ar@{-}[ur]\ar@{-}[dr]\ar@[red]@{-}[r]& \circ\ar@[red]@{-}[r] & \circ\ar@[red]@{-}[r] &\cdots\ar@[red]@{-}[r]& \circ\ar@[red]@{-}[r]& \circ\ar@[red]@{-}[r]& \bullet &\overset{N+1}{\circ}\ar@{-}[d]\\
&\bullet\ar@{-}[r]&\bullet\ar@{-}[r]&\cdots\ar@{-}[r]&\bullet\ar@{-}[r]&\bullet\ar@{-}[ur]&&\overset{N-1}{\square}
}\\
\underbrace{\phantom{\xymatrix{&&&&&}}}_{N\text{ times}}\quad\quad\qquad
\end{gathered}
\ee
From the vertical reduction one obtains a quiver $SU(2)$ gauge theory with bifundamental matter, therefore naively one could expect to preserve an overall diagonal $\mathbb Z_2$ action. Also in this case geometry reveals the presence of discrete theta terms: proceeding as above we can cut open this geometry in a way compatible with the ruling associated to the $SU(2)$ quiver description
\be
\begin{gathered}
\xymatrix{
&\bullet\ar@{-}[r]\ar@[red]@{-}[d]&\bullet\\
\bullet\ar@{-}[ur]\ar@{-}[dr]& \circ\ar@[red]@{-}[d]\\
&\bullet\ar@{..}[uur]} 
\end{gathered} +\begin{gathered}\xymatrix{
&\bullet\ar@{-}[r]&\bullet\\
&\circ\ar@[red]@{-}[d]\ar@[red]@{-}[u]&\\
\bullet\ar@{..}[uur]\ar@{-}[r]&\bullet\ar@{..}[uur]
}\end{gathered} \cdots \begin{gathered}\xymatrix{
&\bullet\ar@{-}[r]&\bullet\\
&\circ\ar@[red]@{-}[d]\ar@[red]@{-}[u]&\\
\bullet\ar@{..}[uur]\ar@{-}[r]&\bullet\ar@{..}[uur]
}\end{gathered}+ \begin{gathered}
\xymatrix{
&\bullet\ar@{-}[dr]\ar@[red]@{-}[d]&\\
& \circ\ar@[red]@{-}[d] & \bullet\\
\bullet\ar@{-}[r]\ar@{..}[uur]&\bullet\ar@{-}[ur]&
}
\end{gathered}
\ee
which corresponds to the surface-quiver diagram
\be
\xymatrix{dP_1 \ar@{-}[r]& \mathbb F_0 \ar@{-}[r]& \cdots \ar@{-}[r]& \mathbb F_0\ar@{-}[r]& dP_1}
\ee
and the two $SU(2)$ factors at the end of the quiver tails carry a nontrivial discrete theta term. For this reason the corresponding instantonic BPS states are charged with respect to the center symmetry thus breaking the overall diagonal $\mathbb Z_2$.

\subsection{Dualities and modified prescription for non-isolated toric singularities}
\label{sec:non-isolated-prescription}

The example above corresponds to geometries that have curves of
singularities giving rise to a non abelian $SU(2N-2)^{(0)}$ global
symmetry. The corresponding singularity is non-isolated, and therefore
the prescription we introduced in \S\ref{sec:toric-general} to compute
the defect groups geometrically needs to be slightly
modified. Motivated by the discussion above, it is natural to
conjecture that whenever the geometry has non-compact curves of
singularities corresponding to marked points on an outer edge of the
toric diagram, the only triangles
$T_i=\langle p_{i-1},p_i,p_{i+1}\rangle$ that should be included in
the computation are those where $p_i$ is \emph{not} along an edge of
the toric diagram. We will call the vertices appearing in such
triangles ``good'' outer vertices. Let us consider the above
example: the good outer vertices are in green
\begin{equation}
\begin{gathered}
\xymatrix{
&\textcolor{green}{\bullet}\ar@{-}[r]&\textcolor{green}{\bullet}\ar@{-}[r]&\bullet \ar@{-}[r]&\cdots\ar@{-}[r]&\bullet\ar@{-}[r]&\textcolor{green}{\bullet}\ar@{-}[r]& \textcolor{green}{\bullet}\ar@{-}[dr]&\\
\textcolor{green}{\bullet}\ar@{-}[ur]\ar@{-}[dr]& \circ & \circ&\circ &\cdots& \circ& \circ &  \circ & \textcolor{green}{\bullet}\\
&\textcolor{green}{\bullet}\ar@{-}[r]&\textcolor{green}{\bullet}\ar@{-}[r] \ar@{..}[ull]&\bullet\ar@{-}[r]&\cdots\ar@{-}[r]&\bullet\ar@{-}[r]&\textcolor{green}{\bullet}\ar@{-}[r]&\textcolor{green}{\bullet}\ar@{-}[ur]&
}
\end{gathered}
\end{equation}
and it is clear that there is at least one outer triangle with minimal area (dashed line above), thus leading to a trivial defect group.

\medskip

\begin{figure}
\begin{center}
\begin{tabular}{cc}
$
\begin{gathered}
\xymatrix@=0.7em{
&&\textcolor{green}{\bullet}\ar@{-}[r]&\textcolor{green}{\bullet}\ar@{-}[r]\ar@{..}[ddll]&\bullet\ar@{-}[r]&\cdots\ar@{-}[r]&\bullet\ar@{-}[r]&\textcolor{green}{\bullet}\ar@{-}[r]& \textcolor{green}{\bullet}\ar@{-}[ddl]&\\
&& \circ & \circ &\cdots& \circ&\circ& \circ\\
&\textcolor{green}{\bullet}\ar@{-}[uur]\ar@{-}[r]&\textcolor{green}{\bullet}\ar@{-}[r]&\bullet\ar@{-}[r]&\cdots\ar@{-}[r]&\bullet\ar@{-}[r]&\textcolor{green}{\bullet}\ar@{-}[r]&\textcolor{green}{\bullet}&
}
\end{gathered}
$&$
\begin{gathered}
\xymatrix@=0.7em{
&&\textcolor{green}{\bullet}\ar@{-}[r]\ar@{..}[dd]&\textcolor{green}{\bullet}\ar@{-}[r]&\bullet\ar@{-}[r]&\cdots\ar@{-}[r]&\bullet\ar@{-}[r]&\textcolor{green}{\bullet}\ar@{-}[r]& \textcolor{green}{\bullet}\ar@{-}[ddl]&\\
&& \circ & \circ &\cdots& \circ&\circ& \circ\\
&\textcolor{green}{\bullet}\ar@{-}[uur]\ar@{-}[r]&\textcolor{green}{\bullet}\ar@{-}[r]&\bullet\ar@{-}[r]&\cdots\ar@{-}[r]&\bullet\ar@{-}[r]&\textcolor{green}{\bullet}\ar@{-}[r]&\textcolor{green}{\bullet}&
}
\end{gathered}
$\\
\\
\end{tabular}
\end{center}
\caption{Computation of the defect group for the $(SU(2)_0)^N$ theory: all the allowed triangles have twice the minimal area. Here we draw two of the four allowed triangles - the other two are the symmetric ones, on the other side of the figure.}\label{fig:ata}
\end{figure}
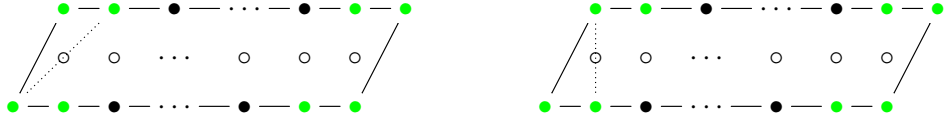

As a further consistency check let us consider the theory 
\be\label{eq:theSU2_0quiver}
\underbrace{SU(2)_0 \times \cdots \times SU(2)_0}_{N \text{ times}}
\ee
is realized as a $\TX$ where $\sing$ is the toric CY singularity with toric diagram
\be
\begin{gathered}
\xymatrix{
&&\bullet\ar@{-}[r]&\bullet\ar@{-}[r]&\cdots\ar@{-}[r]&\bullet\ar@{-}[r]&\bullet\ar@{-}[r]& \bullet\ar@{-}[ddl]&\\
&& \circ & \circ &\cdots& \circ& \circ\\
&\bullet\ar@{-}[uur]\ar@{-}[r]&\bullet\ar@{-}[r]&\bullet\ar@{-}[r]&\cdots\ar@{-}[r]&\bullet\ar@{-}[r]&\bullet&
}\\
\underbrace{\phantom{\xymatrix{&&&&&}}}_{N\text{ times}}
\end{gathered}
\ee
Clearly this does not admit a horizontal reduction, however the corresponding vertical reduction gives rise to the quiver theory in equation \eqref{eq:theSU2_0quiver}. According to the gauge theory analysis this geometry should correspond to a model that has a nontrivial $\mathbb Z_2^{(1)} \times \mathbb Z_2^{(2)}$ discrete higher symmetry group: the instantons are neutral with respect to the diagonal $\mathbb Z_2$ center simmetry, which is unbroken. The fact that all discrete theta terms are zero corresponds to the fact that this geometry has the surface-quiver diagram
\be
\xymatrix{\mathbb F_0 \ar@{-}[r]&\mathbb F_0 \ar@{-}[r] \cdots \ar@{-}[r]& \mathbb F_0}
\ee
Coloring the good outer vertices according to our modified prescription we obtain the graph in figure \ref{fig:ata}: it is clear that the corresponding outer triangles obtained by connecting 3 green adjacent vertices all have area 1. According to our prescription the corresponding defect group is $\mathbb Z_2^{(1)}\times \bZ_2^{(2)}$, which clearly matches with the field theory prediction.

\medskip

As a further consistency, let us consider another variation on the theme above, the theory
\be
\underbrace{SU(K)_0 \times \cdots \times SU(K)_0}_{N \text{ times}}
\ee
is expected, from a similar token, to have a defect group $\mathbb Z_K^{(1)} \times \mathbb Z_K^{(2)}$, arising from the diagonal $\mathbb Z_K$ center symmetry which leaves the bifundamentals invariant. The corresponding colored toric diagram is a trapezium of height $K$:
\be
\begin{gathered}
\xymatrix@=0.8em{
&&\textcolor{green}{\bullet}\ar@{-}[r]&\textcolor{green}{\bullet}\ar@{-}[r]&\bullet\ar@{-}[r]&\cdots\ar@{-}[r]&\bullet\ar@{-}[r]&\textcolor{green}{\bullet}\ar@{-}[r]& \textcolor{green}{\bullet}\ar@{-}[ddddl]&\\
&& \circ & \circ &\cdots& \circ&\circ& \circ\\
&& \vdots & &\cdots& &\vdots& \vdots\\
&& \circ & \circ &\cdots& \circ&\circ& \circ\\
&\textcolor{green}{\bullet}\ar@{-}[uuuur]\ar@{-}[r]&\textcolor{green}{\bullet}\ar@{-}[r]&\bullet\ar@{-}[r]&\cdots\ar@{-}[r]&\bullet\ar@{-}[r]&\textcolor{green}{\bullet}\ar@{-}[r]&\textcolor{green}{\bullet}&
}\\
\underbrace{\phantom{\xymatrix{&&&&}}}_{N\text{ times}}
\end{gathered}
\ee
The corresponding allowed triangles all have area $K/2$, and therefore the corresponding defect group is indeed $\mathbb Z_K^{(1)} \times \mathbb Z_K^{(2)}$ also from geometry, thus matching the gauge theory expectation.

\paragraph{Geometric interpretation.} It is not very difficult to
argue that the modified prescription that we have given is indeed the
correct one geometrically. Recall that the key observation in the
analysis in \cite{Garcia-Etxebarria:2016bpb}, reviewed in
\S\ref{sec:toric-general}, is that for the purposes of computing the
torsion of $Y_5$ one can replace $Y_5$ by a chain of lens spaces
$L_{n_i}$, connected on their non-trivial one-cycle. Each lens space
is associated with a triangle $T_i$ formed by the three consecutive
boundary vertices $\{p_{i-1},p_i,p_{n+1}\}$, and $n_i$, the degree of
the torsion group of the lens space, is twice the area of $T_i$.

\begin{figure}
\begin{center}
  \begin{minipage}{2cm}
    $\begin{gathered}\xymatrix@=0.5em{&\bullet_{v-1}\ar@{-}[r]\ar@{-}[dl]&\cdots\\ \bullet_v\ar@{-}[d]&&&&\vdots\\ \bullet_1\ar@{-}[r]&\bullet_2\ar@{-}[r]&\bullet_3\ar@{-}[r]&\bullet_4\ar@{-}[ur]}\end{gathered}$
    \vspace{3cm}
  \end{minipage}
  \hspace{3cm} \includegraphics[scale=0.6]{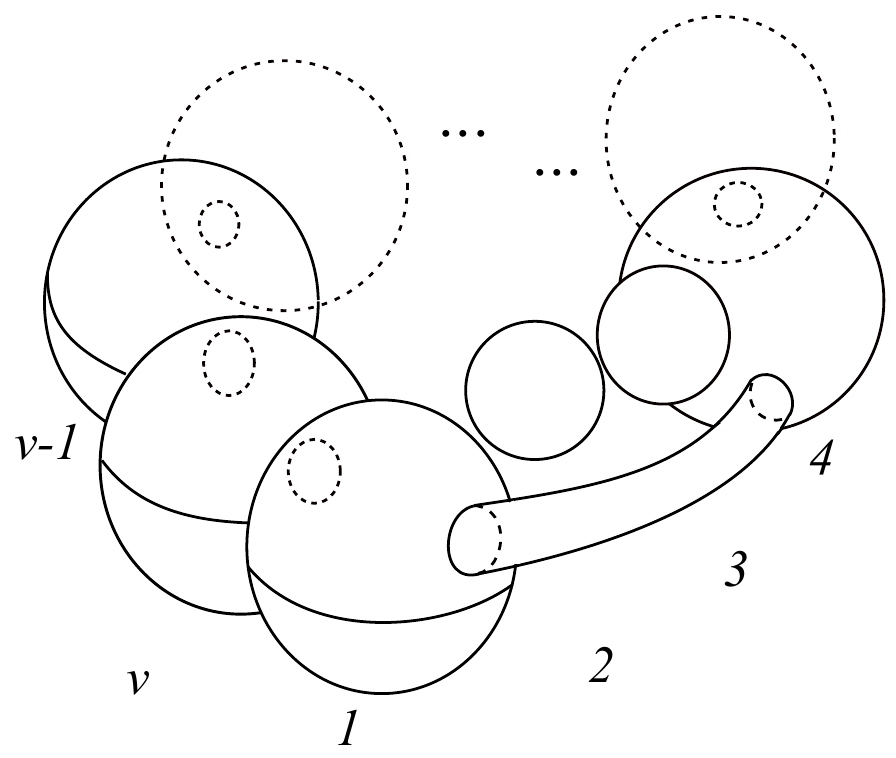}
  \vspace{-3cm}
\end{center}
\caption{Schematic topology of $B_3^L$}\label{fig:modifiedballs}
\end{figure}

Whenever we have a point $p_i$ along an edge, the triangle $T_i$ will
have zero area, so this is formally $L_0$. If we define the lens space
$L_n$ as a circle fibration over $S^2$ of degree $n$, then
$L_0\cong S^2\times S^1$. This is the right answer from the toric
geometry: whenever we have points along an edge, upon crepant
resolution the local geometry of the $T^2$ fiber considered in
\cite{Garcia-Etxebarria:2016bpb} around the point is that of
$S^2\times S^1$ --- see figure \ref{fig:modifiedballs}. Connecting the $S^1$ factor to torsion cycles on
either side is homotopically equivalent to connecting the torsion
cycles directly, so effectively one can ignore the $L_0$ factors,
which is the prescription we used above. Alternatively, we can include
these triangles, and simply state that the prescription is still as
in~\eqref{eq:toric-torsion}:
\begin{equation}
  \sZ = \bZ_{\gcd(n_1,\ldots,n_v)}
\end{equation}
taking into account that $\gcd(0,\ldots)=\gcd(\ldots)$.

\medskip

In the next section, exploiting this prescription, we analyze several
strongly coupled examples and determine the corresponding global
structure. Having done that we give more applications in the context
of other 5d ``dualities'' involving strongly coupled conformal matter
in Section \ref{sec:5dstronglydualz}.

\section{Non-Lagrangian examples}\label{sec:nonlagra}

In this section we exploit the formalism developed above to study the defect group and global structures of 5d SCFTs without a 5d gauge theory phase.

\subsection{The global structure of the 5d $E_0$ SCFT}
The simplest example of model that does not admit any gauge theory phase is given by the 5d $E_0$ SCFT associated to the toric canonical singularity $\mathbb C^3 / \mathbb Z_3$, with toric diagram
\be\label{eq:AD3}
\xymatrix{
&\bullet&\\
&\circ&\bullet\ar@{-}[ul]\\
\bullet\ar@{-}[uur]\ar@{-}[urr]
}
\ee
The naive defect group for the $E_0$ theory is\footnote{ Here we are ignoring the $\mathbb Z_3^{(0)}$ symmetry coming from the isometries --- we thank Kantaro Ohmori for pointing this out. In this particular case it is easy to see that it acts trivially on the fluxes, and therefore it is a global symmetry for all the theories in this class.\label{foot:note}}
\be
\mathbb Z_3^{(1)} \times \mathbb Z_3^{(2)}
\ee
since in the toric diagram \eqref{eq:AD3} there is a single outer triangle with area 3/2. This can be understood both from the perspective of the charged defects and from the perspective of the corresponding charge operators. 

\medskip

The model has a non-trivial BPS string supported on the exceptional $\mathbb P^2$ which arises by resolving the singularity, as well as non-trivial BPS state corresponding to the only compact curve in the geometry $\mathcal C$. This BPS string has a nontrivial Dirac pairing with the corresponding BPS particles:
\be
\mathsf{D} \equiv \mathcal C_x \cdot \mathbf{E} = -3\,.
\ee
therefore the defect group for this model is 
\be
\mathbb D = (\mathbb Z_3)^{(1)}_{M2} \oplus  (\mathbb Z_3^{(2)})_{M5}
\ee
Let's now study this model from the perspective of the flux operators
as well. The analysis of this model proceeds similarly to the
Lagrangian cases we discussed in \S\ref{sec:su(p)_q}, but now we
have no gauge theoretical interpretation of the result. Nevertheless,
if we had one we would expect the different global forms to be
associated to possible choices of flux backgrounds in
\begin{equation}
  \label{eq:gauge-group-5d}
  H^2(\cM_5;\sZ)_e\times H^3(\cM_5;\sZ)_m\, .
\end{equation}
for some $\sZ$ playing the role of the ``center of the gauge
group''. We do obtain this structure from the M-theory construction:
we now have
\begin{equation}
  H^\bullet(S^5/\bZ_3) = \{\bZ, 0, \bZ_3, 0, \bZ_3, \bZ\}\, .
\end{equation}
Using the Künneth formula as in \S\ref{sec:5d-M-theory} we conclude that
$$
\begin{aligned}
&\text{Tor } H^4(\cM_{5}\times (S^5/\bZ_3)) = \Big(H^2(\cM_5)\otimes  \text{Tor } H^2(S^5/\bZ_3)\Big) \oplus \Big(H^0(\cM_5)\otimes  \text{Tor } H^4(S^5/\bZ_3)\Big)\\
&\text{Tor } H^7(\cM_{5}\times (S^5/\bZ_3)) = \Big(H^3(\cM_5)\otimes  \text{Tor } H^4(S^5/\bZ_3)\Big) \oplus \Big(H^5(\cM_5)\otimes  \text{Tor } H^2(S^5/\bZ_3)\Big)
\end{aligned}
$$
We see that we have room for $(-1)$-form and $4$-form background fluxes. We plan to investigate these in the future, for now we focus on the fluxes for the 1-form and the 2-form higher symmetries for this model.

\medskip

The charge operators for the higher 1- and 2- form symmetries of this theory are represented by the decomposition of the M-theory fluxes above. Notice that the M2 flux $\Phi_a$ is parametrized by a class $a\in H^2(\cM_5;\bZ_3)$, while the the M5 flux $\Psi_b$ is parametrized by a class $b\in H^3(\cM_5;\bZ_3)$. We can decompose $a = \omega_2 \otimes \mathsf{a}$ where $\mathsf{a} \in \text{Tor } H^2(S^5/\bZ_3)$ and $\omega_2 \in  H^2(\cM_5)$, and  $b = \omega_3 \otimes \mathsf{b}$ where $\mathsf{b} \in \text{Tor } H^4(S^5/\bZ_3)$ and $\omega_3 \in  H^3(\cM_5)$, then
\be
\Phi_a \Psi_b = e^{2\pi i \mathsf{L}(a,b)}\Psi_b \Phi_a
\ee
where
\be
\mathsf{L}(a,b) \equiv \ell(\mathsf{a},\mathsf{b}) \left( \int_{\cM_5} \omega_2\wedge \omega_3\right)\,.
\ee
The second term in parenthesis gives the linking for the support of the two operators along $\cM_5$, while the first term is given by the linking pairing
\begin{equation}
 \ell : H^2(S^5/\bZ_3)\times H^4(S^5/\bZ_3)\to U(1)\,.
\end{equation}
The only missing piece in order to show that this does indeed lead to a choice of global
structure for this theory is showing that this linking pairing is nontrivial -- but it necessarily is non-trivial due to being perfect. Indeed, the general computation in appendix~\ref{app:E0rankp} gives
\be
\ell(1,1) = -{1\over 3}
\ee
and therefore the possible choices of global structure for this model are the same as those of the $\mathfrak{su}(3)$ theory in 4d \cite{Aharony:2013hda}.

\medskip

As pointed out in \cite{Aharony:1997bh}, this model is the 5d analogue of an Argyres-Douglas theory \cite{Argyres:1995jj}, and indeed it arises along the Coulomb branch of the 5d $\cN=1$ gauge $SU(2)_\pi$ gauge theory, precisely by suitable tuning, as manifest from the corresponding toric diagram --- recall that $SU(2)_\pi$ corresponds to the CY threefold obtained from a  $dP_1$ base:
\be
\begin{gathered}
\xymatrix{
&\bullet&\\
&\circ&\bullet\ar@{-}[ul]\\
\bullet\ar@{-}[uur]\ar@{..}[urr]\ar@{-}[r]&\bullet\ar@{-}[ur]
}
\end{gathered} \quad\xrightarrow{\text{decoupling}} \begin{gathered}\xymatrix{
&\bullet&\\
&\circ&\bullet\ar@{-}[ul]\\
\bullet\ar@{-}[uur]\ar@{-}[urr]
}\end{gathered}
\ee
It is interesting to remark that, while $SU(2)_\pi$ does not have nontrivial higher form symmetries, upon decoupling these symmetry emerge. Clearly the particles we are decoupling are charged with respect to this discrete symmetry and thus are breaking it explicitly in the $SU(2)_\pi$ theory. The decoupling is achieved starting from an $SU(2)_\pi$ gauge theory phase by a flop transition involving a massless BPS instanton. Naively one would expect a $\mathbb Z_2$ symmetry, however along this transition the symmetry enhances.

\subsection{The higher rank 5d $E_0$ theories}

The discussion in the previous section carries over word for word for the class of higher rank 5d $E_0$ Argyres-Douglas like theories, which correspond to the M-theory singularities $\bC^3/\bZ_{2p+1}$. These models can be obtained by deforming $SU(p)_1$ as follows \cite{Aharony:1997bh}
\be
\begin{gathered}
\xymatrix@=0.8em{
&\bullet&\\
&\vdots&\\
&\circ&\\
&\circ&\bullet\ar@{-}[uuul]\\
\bullet\ar@{-}[uuuur]\ar@{..}[urr]\ar@{-}[r]&\bullet\ar@{-}[ur]
}
\end{gathered} \quad\xrightarrow{\text{decoupling}} \begin{gathered}\xymatrix@=0.8em{
&\bullet&\\
&\vdots&\\
&\circ&\\
&\circ&\bullet\ar@{-}[uuul]\\
\bullet\ar@{-}[urr]\ar@{-}[uuuur]
}\end{gathered}
\ee
The gauge theory we begin with has trivial defect group, while the higher rank $E_0$ theory has defect group\footnote{ As in footnote \ref{foot:note} here we are ignoring the $\mathbb Z_p^{(0)}$ symmetry coming from the isometries. The same remark applies.} $$\mathbb Z_{2p+1}^{(1)} \times \mathbb Z_{2p+1}^{(2)}\,.$$ An enhancement similar to the one observed in the previous section is in place.  As we shall see below this enhancement is a feature for many descendants of the trinionic 5d SCFTs as well.

\medskip

The global structure in this case is obtained by the same procedure. We have
\be
H^\bullet(S^5/\mathbb Z_{2p+1}) = \{\bZ, 0, \bZ_{2p+1}, 0, \bZ_{2p+1}, \bZ\}\, .
\ee
and therefore, proceeding as in the previous section we obtain
\begin{align}
 \text{Tor } H^4(\cM_{5}\times (S^5/\bZ_3)) &= H^2(\cM_5;\bZ_{2p+1})\oplus H^0(\cM_5;\bZ_{2p+1})\\
 \text{Tor }  H^7(\cM_5\times (S^5/\bZ_3)) &=  H^3(\cM_5;\bZ_{2p+1}) \oplus H^5(\cM_5;\bZ_{2p+1})\, .
\end{align}
Ignoring the (-1)-form and the 4-form symmetries, we focus on the M2 flux $\Phi_a$ parametrized by a class $a\in H^2(\cM_5;\bZ_{2p+1})$, and on the M5 flux $\Psi_b$ parametrized by a class $b\in H^3(\cM_5;\bZ_{2p+1})$. We have that 
\be
\Phi_a \Psi_b = e^{2\pi i \mathsf{L}(a,b)}\Psi_b \Phi_a
\ee
and decomposing $a = \omega_2 \otimes \mathsf{a}$ and  $b = \omega_3 \otimes \mathsf{b}$
\be
\mathsf{L}(a,b) \equiv \ell(\mathsf{a},\mathsf{b}) \left( \int_{\cM_5} \omega_2\wedge \omega_3\right)\,.
\ee
where $\ell(\mathsf{a},\mathsf{b})$ is the linking pairing (see appendix \ref{app:E0rankp})
\be
\ell(1,1)= -{1\over 2p+1}\,.
\ee
we see that this linking pairing has the same structure (up to a sign) as the pairing for the 4d $\mathfrak{su}(2p+1)$ algebras, and therefore the rank $p$ $E_0$ theories have the same choices of global structures! These global structures have been worked out in \cite{Aharony:2013hda}, to which we refer the interested readers.

\subsection{The 5d trinions and their descendants}

Consider the singularity corresponding to the 5d $T_N$ theory \cite{Benini:2009gi} \be\mathcal V_N \equiv {\mathbb C^3 \over \bZ_N \times \bZ_N}\,.\ee The corresponding SCFT has 0-form global symmetry $(SU(N)^{(0)})^3$, which is manifest from the fact that there are three lines of singularities $\mathbb C^2 / \bZ_N \times \bC$ meeting at the origin. The corresponding colored toric diagram is
\be
\begin{gathered}
\xymatrix@=0.8em{
\textcolor{green}{\bullet}\ar@{-}[d]\ar@{-}[dr]\\
\textcolor{green}{\bullet}\ar@{-}[d]&\textcolor{green}{\bullet}\ar@{-}[dr]\\
\bullet\ar@{-}[d]&&\bullet\ar@{-}[dr]\\
\vdots\ar@{-}[d]&&&\ar@{-}[dr]\ddots\\
\bullet\ar@{-}[d]&&&&\bullet\ar@{-}[dr]\\
\textcolor{green}{\bullet}\ar@{-}[d]&&&&&\textcolor{green}{\bullet}\ar@{-}[dr]\\
\textcolor{green}{\bullet}\ar@{-}[r]&\textcolor{green}{\bullet}\ar@{-}[r]&\bullet\ar@{-}[r]&\cdots\ar@{-}[r]&\bullet\ar@{-}[r]&\textcolor{green}{\bullet}\ar@{-}[r]&\textcolor{green}{\bullet}
}
\end{gathered}
\ee
where each edge has $N+1$ bullets. It is manifest that there are triangles of minimal area above, therefore this theory has a trivial defect group. This fact can be understood easily because the 5d trinion has a gauge theory phase \cite{Bergman:2014kza,Hayashi:2014hfa}
\be
\xymatrix{\overset{N}{\square}\ar@{-}[r]&\overset{N-1}{\circ}\ar@{-}[r]&\overset{N-2}{\circ}\ar@{-}[r]&\cdots\ar@{-}[r]&\overset{3}{\circ}\ar@{-}[r]&\overset{2}{\circ}\ar@{-}[r]&\overset{2}{\square}}
\ee
and the presence of matter in the fundamental is breaking the diagonal center symmetry.

\medskip

The $\mathcal T_N$ theories have several descendant theories obtained by decoupling in \cite{Eckhard:2020jyr}. As descendant theories are determined by decoupling, these correspond to those convex toric diagrams that embed in the ones associated to the $\mathcal T_N$ theories. In particular notice that all $\mathcal T_K$ theories with $K<N$ are always descendants of $\mathcal T_N$. Using our method is rather easy to identify graphically examples in this class that have non-trivial higher form symmetries by applying the following two criteria 
\begin{enumerate}
\item \textit{Non-Lagrangian}: the descendant geometries with a larger defect group typically do not admit a quiver gauge theory phase, which entails some of the horizontal, vertical and diagonal rulings must be obstructed; 
\item \textit{High gcd}: the outer green triangles should have areas with non-trivial greater common divisor
\end{enumerate}
Some examples of descendant theories with a nontrivial higher form symmetry are listed in figure \ref{fig:uradimerda}. A systematic analysis of the remaining cases is beyond the scope of the present note.

\begin{figure}
\begin{center}
\begin{tabular}{cc}
\phantom{\Big|}descendant & defect group \\
\hline
&\\
$
\begin{gathered}
\xymatrix@=0.3em{
\bullet\ar@{-}[d]\ar@{-}[dr]\\
\textcolor{red}{\bullet}\ar@[red]@{-}[d]\ar@[red]@{-}[dddddrrrrrr]&\bullet\ar@{-}[dr]\\
\textcolor{red}{\bullet}\ar@[red]@{-}[d]&&\bullet\ar@{-}[dr]\\
\vdots\ar@[red]@{-}[d]&&&\ar@{-}[dr]\ddots\\
\textcolor{red}{\bullet}\ar@[red]@{-}[d]&&&&\bullet\ar@{-}[dr]\\
\textcolor{red}{\bullet}\ar@{-}[d]\ar@[red]@{-}[drrrrrr]&&&&&\bullet\ar@{-}[dr]\\
\bullet\ar@{-}[r]&\bullet\ar@{-}[r]&\bullet\ar@{-}[r]&\cdots\ar@{-}[r]&\bullet\ar@{-}[r]&\bullet\ar@{-}[r]&\textcolor{red}{\bullet}
}
\end{gathered}
$& $\bZ_{N}^{(1)} \times \bZ_{N}^{(2)}$\\
&\\
\hline
&\\
$
\begin{gathered}
\xymatrix@=0.3em{
\bullet\ar@{-}[d]\ar@{-}[dr]\\
\bullet\ar@{-}[d]&\textcolor{red}{\bullet}\ar@[red]@{-}[dddddrrrr]\ar@[red]@{-}[ddddl]\ar@{-}[dr]\\
\bullet\ar@{-}[d]&&\bullet\ar@{-}[dr]\\
\vdots\ar@{-}[d]&&&\ar@{-}[dr]\ddots\\
\bullet\ar@{-}[d]&&&&\bullet\ar@{-}[dr]\\
\textcolor{red}{\bullet}\ar@{-}[d]\ar@[red]@{-}[drrrrr]&&&&&\bullet\ar@{-}[dr]\\
\bullet\ar@{-}[r]&\bullet\ar@{-}[r]&\bullet\ar@{-}[r]&\cdots\ar@{-}[r]&\bullet\ar@{-}[r]&\textcolor{red}{\bullet}\ar@{-}[r]&\bullet
}
\end{gathered}
$& $\bZ_{N^2-3(N-1)}^{(1)} \times \bZ_{N^2-3(N-1)}^{(2)}$\\
&\\
\end{tabular}
\end{center}

\caption{Examples of $\mathcal T_N$ descendants \cite{Eckhard:2020jyr} with a nontrivial center symmetry. The one on the bottom is known as the $B_N$ theory \cite{Eckhard:2020jyr}.}\label{fig:uradimerda}
\end{figure}

\subsection{More about 5d ``dualities''}\label{sec:5dstronglydualz}
It is interesting to remark that the 5d quiver theory $(SU(K)_0)^N$ has 5d ``duals'' that are realized via quiver theories with strongly coupled trinionic 5d conformal matter. As a concrete example, consider the theory $(SU(4)_0)^6$, which belongs to the family of models we have analyzed in the previous paragraph. Considering the following ruling of this geometry:
\be
\begin{gathered}
\xymatrix@=0.8em{
&&\bullet\ar@{-}[r]&\bullet\ar@{-}[r]&\bullet\ar@{-}[r]&\bullet\ar@{-}[r]&\bullet\ar@{-}[r]&\bullet\ar@{-}[r]& \bullet\ar@{-}[ddddl]&\\
&& \circ & \circ &\circ\ar@[red]@{-}[ur]& \circ\ar@[red]@{-}[ur]&\circ\ar@[red]@{-}[ur]& \circ\ar@[red]@{-}[ur]\\
&& \circ & \circ\ar@[red]@{-}[ur] &\circ\ar@[red]@{-}[ur]& \circ\ar@[red]@{-}[ur]&\circ\ar@[red]@{-}[ur]& \circ\\
&& \circ \ar@[red]@{-}[ur]& \circ\ar@[red]@{-}[ur] &\circ\ar@[red]@{-}[ur]& \circ\ar@[red]@{-}[ur]&\circ& \circ\\
&\bullet\ar@{-}[uuuur]\ar@{-}[r]\ar@[red]@{-}[ur]&\bullet\ar@{-}[r]\ar@[red]@{-}[ur]&\bullet\ar@{-}[r]\ar@[red]@{-}[ur]&\bullet\ar@{-}[r]\ar@[red]@{-}[ur]&\bullet\ar@{-}[r]&\bullet\ar@{-}[r]&\bullet&
}\end{gathered}
\ee
It is manifest that this model admits a different quiver description, namely
\be
\xymatrix{\overset{\TX}{\diamond} \ar@{-}[r]&\overset{4_0}{\circ}\ar@{-}[r]&\overset{4_0}{\circ}\ar@{-}[r]&\overset{4_0}{\circ}\ar@{-}[r]&\overset{4_0}{\circ} \ar@{-}[r] &\overset{\TX}{\diamond}}
\ee
where, as usual, the edges denote bifundamentals, while nodes $\overset{N_k}{\circ}$ correspond to a gauge subsector $SU(N)_k$, but we have introduced another kind of ``meta-node,'' $\diamond$, which corresponds to a strongly coupled sector. Edges connecting the meta-nodes to a gauge node indicate the gauging of a zero-form global symmetry. In the context of this example, we have two meta-nodes associated to two copies of the same $\TX$ theory, where $\sing$ is the toric canonical singularity associated to the toric diagram
\be
\begin{gathered}
\xymatrix@=0.8em{
&&\bullet\ar@{-}[r]&\bullet\ar@{-}[r]&\bullet\ar@{-}[r]& \bullet\ar@{-}[ddddllll]&\\
&& \circ & \circ &\bullet\\
&& \circ & \bullet\\
&& \bullet\\
&\bullet\ar@{-}[uuuur]
}\end{gathered}
\ee

This model is among the descendant theories of the 5d $\mathcal T_4$
trinion and clearly has a global symmetry
$SU(4)^{(0)} \times SU(3)^{(0)}$, and no 1-form symmetries. From the
duality, it is manifest that the operators in this theory must be
invariant under the $\mathbb Z_4$ center symmetry of $SU(4)$, which
survives the gauging, and provides the necessary structure for the
gauged theory to have the $\bZ_4$ center symmetry.

\bigskip

We can also consider gauging this flavor symmetry only, thus producing yet another 5d SCFT corresponding to the toric diagram below
\be
\begin{gathered}
\xymatrix@=0.8em{
&&\textcolor{green}{\bullet}\ar@{-}[r]&\textcolor{green}{\bullet}\ar@{-}[r]&\bullet\ar@{-}[r]& \textcolor{green}{\bullet}\ar@[red]@{-}[dl]\ar@{-}[r]&\textcolor{green}{\bullet}\\
&& \circ & \circ &\textcolor{red}{\bullet}\ar@[red]@{-}[dl]&\\
&& \circ & \textcolor{red}{\bullet}\ar@[red]@{-}[dl]&\\
&& \textcolor{red}{\bullet}\ar@[red]@{-}[dl]&\\
&\textcolor{green}{\bullet}\ar@{-}[uuuur]\ar@{-}[uuuurrrrr]
}\end{gathered}
\ee
here we have marked in red the nodes corresponding to the $SU(4)$ gauge symmetry, and in green the nodes which enters the computation of the defect groups. The resulting theory has quiver
\be
\xymatrix{\overset{\TX}{\diamond} \ar@{-}[r]&\overset{4_0}{\circ}}
\ee
and defect group $\mathbb Z_4^{(1)} \times \mathbb Z_4^{(2)}$. Notice that we can introduce a nontrivial CS level for the $SU(4)$ gauge theory, leading to the gauge theories
\be
\xymatrix{\overset{\TX}{\diamond} \ar@{-}[r]&\overset{4_p}{\circ}}
\ee
with defect groups $\mathbb Z_{\text{gcd}(4,p)}^{(1)} \times \mathbb Z_{\text{gcd}(4,p)}^{(2)}$, corresponding to the following toric diagrams
\be
\begin{gathered}
\begin{gathered}
\xymatrix@=0.7em{
&&\textcolor{green}{\bullet}\ar@{-}[r]&\textcolor{green}{\bullet}\ar@{-}[r]&\textcolor{green}{\bullet}\ar@{-}[r]& \textcolor{green}{\bullet}\ar@[red]@{-}[dl]\ar@{-}[d]&\\
&& \circ & \circ &\textcolor{red}{\bullet}\ar@[red]@{-}[dl]&\textcolor{green}{\bullet}\\
&& \circ & \textcolor{red}{\bullet}\ar@[red]@{-}[dl]&\\
&& \textcolor{red}{\bullet}\ar@[red]@{-}[dl]&\\
&\textcolor{green}{\bullet}\ar@{-}[uuuur]\ar@{-}[uuurrrr]
}
\end{gathered}\,\begin{gathered}
\xymatrix@=0.7em{
&&\textcolor{green}{\bullet}\ar@{-}[r]&\textcolor{green}{\bullet}\ar@{-}[r]&\textcolor{green}{\bullet}\ar@{-}[r]& \textcolor{green}{\bullet}\ar@[red]@{-}[dl]\ar@{-}[ddl]&\\
&& \circ & \circ &\textcolor{red}{\bullet}\ar@[red]@{-}[dl]&\\
&& \circ & \textcolor{red}{\bullet}\ar@[red]@{-}[dl]&\textcolor{green}{\bullet}\\
&& \textcolor{red}{\bullet}\ar@[red]@{-}[dl]&\\
&\textcolor{green}{\bullet}\ar@{-}[uuuur]\ar@{-}[uurrr]
}
\end{gathered}\,\begin{gathered}
\xymatrix@=0.7em{
&&\textcolor{green}{\bullet}\ar@{-}[r]&\textcolor{green}{\bullet}\ar@{-}[r]&\textcolor{green}{\bullet}\ar@{-}[r]& \textcolor{green}{\bullet}\ar@[red]@{-}[dl]\ar@{-}[dddll]&\\
&& \circ & \circ &\textcolor{red}{\bullet}\ar@[red]@{-}[dl]&\\
&& \circ & \textcolor{red}{\bullet}\ar@[red]@{-}[dl]&\\
&& \textcolor{red}{\bullet}\ar@[red]@{-}[dl]&\textcolor{green}{\bullet}\\
&\textcolor{green}{\bullet}\ar@{-}[uuuur]\ar@{-}[urr]
}
\end{gathered}
\end{gathered}
\ee In these examples, the center of the gauge group is broken
(partially, for $p=2$) by the charges of the corresponding
instantons. It is interesting to remark that the second model has a
different 5d ``dual'' description \be
\begin{gathered}
\xymatrix@=0.7em{
&&\textcolor{green}{\bullet}\ar@{-}[r]\ar@[red]@{-}[dr]&\textcolor{green}{\bullet}\ar@{-}[r]&\textcolor{green}{\bullet}\ar@{-}[r]& \textcolor{green}{\bullet}\ar@{-}[ddl]&\\
&& \circ & \textcolor{red}{\bullet}\ar@[red]@{-}[dr] &\bullet&\\
&& \circ & \bullet&\textcolor{green}{\bullet}\\
&& \bullet&\\
&\textcolor{green}{\bullet}\ar@{-}[uuuur]\ar@{-}[uurrr]
}
\end{gathered}
\ee
which has a generalized quiver
\be
\xymatrix{\overset{\mathcal T_1}{\diamond} \ar@{-}[r]&\overset{2_0}{\circ}\ar@{-}[r]&\overset{\mathcal T_2}{\diamond}}
\ee
where we see another manifestation of the fact that the descendant of the 5d trinions respect the center of the gauge group upon gauging, which in this case gives a defect group that is precisely $\mathbb Z_2^{(1)} \times \mathbb Z_2^{(2)} $.

\medskip

These examples (which easily generalize) illustrate another
application of 5d ``dualities'' to understand the structure of the
operators of strongly coupled 5d SCFTs. (A somewhat analogous
argument, applied in reverse, was used in \cite{Tachikawa:2013hya} to
learn about the 1-form symmetries of the $T_N$ theories.)

\section{Global structure of 4d $\cN=1$ theories and M-theory on $G_2$ spaces}\label{sec:4d}

In this section we give an appetizer about the application of our method in the context of M-theory on spaces with $G_2$ holonomy. More precisely we consider here a geometric engineering on a space
\be
\mathcal \cM_4 \times \mathcal V_7
\ee
where $\mathcal V_7$ is a local $G_2$ space. For the applications we have in mind,
\be
\mathcal V_7  = \mathcal C (Y_6)
\ee
where $Y_6$ is a nearly K\"ahler manifold or an orbifold thereof  --- see e.g. \cite{10.2307/24906437}. It is well-known that M-theory on a $G_2$ cone gives rise to four-dimensional quantum field theories with $\cN=1$ supersymmetry (see e.g. \cite{Acharya:2004qe} and references therein for a nice review). Our task in this section is to determine the global structure of one such theory.

\medskip

Repeating the same arguments as above we expect this is determined from the structure of flux operators $\Phi_{M2,a}$ with $a\in \Tor H^4(\cM_4 \times Y_6)$ and $\Psi_{M5,b}$ with $b\in \Tor H^7(\cM_4 \times Y_6)$ (measuring torsional M2 and M5 charge, respectively). Again we are interested in studying the structure of
\begin{equation}
  \Phi_{M2,a}\Psi_{M5,b} = e^{2\pi i \sL(a,b)}\Psi_{M5,b}\Phi_{M2,a}
\end{equation}
where
\begin{equation}
  \sL \colon \Tor H^4(\cM_4 \times Y_6) \times \Tor H^7(\cM_4 \times Y_6) \to \bQ/\bZ
\end{equation}
is the linking pairing in $\cM_4 \times Y_6$.

\medskip

Assuming that $\cM_4$ has no torsion we are interested in the cohomology groups
\be
\Tor H^m(\cM_4 \times Y_6) = \sum_{i+j=m} H^i(\cM_4) \otimes \Tor H^j(Y_6)\qquad m = 4,7
\ee
Poincaré duality, together with the universal coefficient theorem,
implies that
\be
\sZ = \Tor  H^2(Y_6) = \Tor  H^5(Y_6) \qquad \tilde{\sZ} =  \Tor H^3(Y_6) = \Tor H^4(Y_6) 
\ee
are the only nonzero torsional parts from the cohomology of $Y_6$, and therefore the relevant parts of $\KM(\cM_4 \times Y_6)$ that can give rise to a nontrivial global structure are (by the universal coefficient theorem)
\be\label{eq:4dglobal}
\begin{aligned}
\Tor H^4(\cM_4 \times Y_6)&= H^2(\cM_4,\sZ)  \oplus H^1(\cM_4,\tilde{\sZ}) \oplus H^0(\cM_4,\tilde{\sZ})\\
\Tor H^7(\cM_4 \times Y_6)&=H^2(\cM_4,\sZ) \oplus H^3(\cM_4,\tilde{\sZ}) \oplus H^4(\cM_4,\tilde{\sZ}) 
\end{aligned}
\ee
where the direct summands above are paired along the vertical direction in the above equation, meaning that for this class of examples we have three distinct commuting Heisenberg algebras acting on $\mathcal H_M(\cM_4 \times Y_6)$. 

\medskip

From the above structure we see that from the first summand there are an electric and a magnetic $\sZ^{(1)}$ form symmetry, from the second summand we have similarly a $\tilde{\sZ}^{(0)}$ form and a $\tilde{\sZ}^{(2)}$ form symmetry, while the last summand corresponds to $\tilde{\sZ}^{(-1)}$ form and a $\tilde{\sZ}^{(3)}$ form symmetries. The defect group for this geometry is 
\be
\mathbb D = \Big(\sZ^{(1)}_{M2} \oplus \sZ^{(1)}_{M5}\Big)\oplus \Big(\tilde{\sZ}^{(0)}_{M2} \oplus \tilde{\sZ}^{(2)}_{M5}\Big) \oplus \Big(\tilde{\sZ}^{(-1)}_{M2} \oplus \tilde{\sZ}^{(3)}_{M5}\Big)
\ee
In this paper we are going to ignore the effects associated to (-1)-form symmetries and the 3-form symmetries.\footnote{ These are somewhat more exotic and we defer their study to the future --- see \cite{Tanizaki:2019rbk} for a reference about these.} 

\medskip

Each of the three summands in parenthesis in \eqref{eq:4dglobal} belongs to a distinct Heisenberg factor, thus signalling a mixed 't Hooft anomaly for the defect group.  The non-commuting 1-form symmetries in this case are the analogues of
the electric and the magnetic 1-form symmetries in Yang-Mills theory
in four-dimensions, so this hardly come as a surprise
\cite{tHooft:1977nqb}.

\medskip

It is interesting to remark however that discrete 0-form charge
operators and 2-form charge operator do not commute. This entails that
we cannot specify a background for the discrete 0-form symmetries of
one such model and a background flux for the 2-form symmetries
simultaneously whenever $\cM_4$ has non-trivial one-cycles and
three-cycles.\footnote{Ultimately this effect is related to the fact
  that scalars are dual to tensors in four dimensions.} An interesting
example of a manifold where this choice needs to be made is
\begin{equation}
  \cM_4 = S^1\times S^3\, .
\end{equation}
We emphasize that the choice that we need to make here is independent
of the choice that we make in the 1-form symmetry sector (note that
generically, $\sZ\neq \tilde\sZ$). So whenever our manifold $\cM_4$
has non-trivial one and two-cycles, there are generically two
independent choices of global structure to be made when computing
partition functions. There are interesting partition functions of this
sort, the simplest is
\begin{equation}
  \cM_4 = T^2 \times S^2\,.
\end{equation}

\subsection{Example: 4d $\cN=1$ SYM}

Here we focus on a simple class of examples of this kind provided by the Bryant-Salamon $G_2$ metric on the spin bundle over $S^3$ (and some of its orbifolds) \cite{bryant1989,Gibbons:1989er}.  This metric has topology $S^3 \times \mathbb R^4$, and we can realize this space as a hyperbolic submanifold of $\mathbb H \times \mathbb H \simeq \bC^2 \times \bC^2$
\be
|q_1|^2 - |q_2|^2 = |z_{1,1}|^2 + |z_{1,2}|^2 -|z_{2,1}|^2  -|z_{2,2}|^2 =  V \qquad V\in\mathbb R_{>0}\,.
\ee
where $z_{i,j}$ are complex coordinates. From this presentation it is clear that we can orbifold this space with a discrete subgroup $\Gamma \subset SU(2)$ acting only on one of the two sets of coordinates, thus obtaining a space that topologically is $S^3 \times \mathbb C^2/\Gamma$. This space is known to geometrically engineer a four-dimensional $\cN=1$ SYM theory in M-theory \cite{Atiyah:2000zz,Acharya:2001hq,Acharya:2001dz}.

\medskip

One feature of this $G_2$ space is that it is a cone $\mathcal C(S^3 \times S^3 /\Gamma)$. It is straightforward to compute 
\be
H^\bullet(S^3 \times S^3 /\Gamma) = \{\bZ,0,\Gamma^{\text{ab}},\bZ\oplus \bZ, 0, \Gamma^{\text{ab}},\bZ\}
\ee
hence for this class of examples the two groups in \eqref{eq:4dglobal} are given by
\be
\sZ = \Gamma^{\text{ab}} = Z(G_\Gamma)  \qquad \tilde{\sZ} = 0\,
\ee
therefore in this specific model there only one Heisenberg algebra. More precisely, for fluxes labelled by $a_i = (\omega \otimes \ell)_i \in H^2(\cM_4)\otimes\sZ$ we have
\be
\Phi_{M2,a_1} \Psi_{M5,a_2} = \exp \left( 2\pi i  \, \mathsf{L}_\Gamma(\ell_1,\ell_2) \int_{\cM_4} \omega_1 \wedge \omega_2\right) \Psi_{M5,a_2} \Phi_{M2,a_1}
\ee 
where $ \sL_\Gamma$ is the same pairing we introduced in section \ref{sec:7d}. We thus reproduce the statement about the mixed 't Hooft anomaly among the $Z(G_\Gamma)^{(1)}_{M2}$ and the $Z(G_\Gamma)^{(1)}_{M5}$ symmetries in the defect group mentioned in the introduction of this paper.

\medskip

More examples of $G_2$ cones are available in the literature. The global structures of these models are interesting and can be analyzed using our method. We will address this question in future work.

\section*{Acknowledgments}

We thank Bobby Acharya, Francesco Benini, Cyril Closset, Stefano
Cremonesi, Lorenzo Foscolo, Fernando Marchesano and Maxim Zabzine for
discussions. This project has received funding from the European
Research Council (ERC) under the European Union's Horizon 2020
research and innovation programme (grant agreement
No. 851931). I.G.-E. is supported in part by STFC consolidated grant
ST/P000371/1. This project was also supported in part by STFC grant
with project reference ST/T506035/1.

\appendix

\section{The Mori cone for $\cC_\bR(Y^{p,q})$}
\label{app:Mori-cone-Y^{p,q}}

In this section we will study in detail the structure of the Mori cone
for the Calabi-Yau cone over $Y^{p,q}$ \cite{Gauntlett:2004yd}, which
we denote by $\cC_\bR(Y^{p,q})$. This Calabi-Yau threefold is toric,
which simplifies the analysis of the relevant geometry. We refer the
reader to~\cite{CLS} for general background on toric geometry and
\cite{Ref,Denef:2008wq,Hori:2003ic} for introductions aimed to
physicists. The computer algebra program \textsc{Sage} contains very
useful implementations of the toric algorithms that we use
\cite{sage}.

Define $l\df p-q$. We can take the points in the toric diagram for
$\cC_\bR(Y^{p,q})$ to be $P_1=(-1,1)$, $P_3=(l,0)$ and $I_i=(0,i),$
$i\in\{0,1,\cdots,p\}$. We choose the triangulation as in
figure~\ref{fig:Y^{p,q}-triangulation}, that is, such that the
$3$-dimensional cones are of the form $(P_1, I_k, I_{k+1})$ and
$(P_3, I_k, I_{k+1})$ with $k\in\{0,1,2,\cdots,p-1\}$.

\begin{table}[t]
\setlength{\tabcolsep}{4pt}
    \centering
    \begin{tabular}{ | m{3.5em} | m{.4cm}| m{.4cm} | m{.8cm} | m{.8cm} |m{0.4cm}| m{0.4cm}| m{0.7cm}| m{1.6cm}| m{1.4cm} |  m{0.4cm}| m{1cm}| m{1.9cm} | } \hline
    Curve&\centering $P_1$ &\centering$P_3$ &\centering $I_0$ &\centering $I_1$&\centering$I_2$ &\centering$\cdots$ &\centering $I_{k-1}$ &\centering $I_k$ &\centering $I_{k+1}$ &\centering $\cdots$ &\centering$I_{p-1}$ &$I_p$ \\ [0.5ex] 
    \hline
    \hline
    $P_1\cdot I_1$&\centering$0$ &\centering$0$&\centering$1$&\centering$-2$&\centering$1$&$\cdots$&\centering$0$&\centering$0$&\centering$0$&$\cdots$&\centering$0$&$0$\\ [0.5ex] 
    \hline
    $\cdots$ &&&&&&&&&&&& \\ [0.5ex] 
    \hline
    $P_1\cdot I_{k}$ &\centering$0$ &\centering $0$&\centering $0$ &\centering$0$&\centering $0$ &\centering$\cdots$ &\centering $1$ &\centering $-2$ &\centering $1$ &\centering$\cdots$ &\centering$0$&$0$\\[0.5ex]  
    \hline 
    $\cdots$ &&&&&&&&&&&& \\ [0.5ex] 
    \hline
    $I_0\cdot I_1$ &\centering$1$ &\centering$1$ &\centering$l-2$ &\centering$-l$ &\centering $0$&\centering$\cdots$ &\centering $0$ &\centering $0$ &\centering $0$ &\centering$\cdots$ &\centering $0$ &$0$ \\ [0.5ex] 
    \hline
    $I_{k}\cdot I_{k+1}$ &\centering $1$ &\centering$1$ &\centering$0$ &\centering $0$ &\centering $0$&\centering$\cdots$ &\centering $0$ &\centering $l-2k-2$ &\centering $-l+2k$ &\centering $\cdots$ &\centering $0$ &$0$\\ [0.5ex] 
    \hline
    $\cdots$ &&&&&&&&&&&&  \\ [0.5ex] 
    \hline
    $I_{p-1}\cdot I_p$ &\centering$1$&\centering$1$&\centering$0$&\centering$0$&$\cdots$&\centering$0$&\centering$0$&\centering$0$&\centering $0$&$\cdots$&\centering$l-2p$&$-l+2(p-1)$\\ [0.5ex] 
    \hline
\end{tabular}
\caption{The intersection numbers of the $(2P-1)$ compact curves
  $P_1\cdot I_i$, $I_i\cdot I_{i-1}$ and the $(P+3)$ points
  $P_1=(-1,1)$, $P_3=(l,0)$ and $I_i=(0,i)$, where
  $i\in\{0,1,\cdots,p\}$, and $k\in\{1,2,\cdots,(p-1)\}$. We have
  omitted the result for the curves $P_3\cdot I_k$ as they give the
  same intersection numbers as $P_1\cdot I_k$ for each fixed $k$.}
\label{t1}
\end{table}

We can construct a (non-minimal) basis of generating curves by taking
intersections of toric divisors. The intersection numbers of the
compact curves constructed in this way and the toric divisors are
given in table~\ref{t1}. The Mori cone is spanned by compact curves
corresponding to $2$-dimensional cones. Thus, the number of the
generators of the Mori cone equals to the number of independent
$2$-cycles. From our discussion in~\eqref{eq:toric-homology} we find
that the number of independent compact $2$-cycles is $p$, so this is
the dimension of the Mori cone. We denote the Mori cone generators
$C_1,\ldots,C_p$. Any two curves are linearly equivalent iff their
intersection with all toric divisors are the same, so the problem of
determining the $C_i$ reduces to finding a basis of linearly
independent rows in table~\ref{t1}. From the table we can deduce the
equivalence relations
\begin{equation}
  P_1\cdot I_k\equiv P_3\cdot I_k,\quad I_{k-1}\cdot I_k-I_k\cdot I_{k+1}\equiv(l-2k)P_1\cdot I_k,
\end{equation}
where $k\in\{1,2,\cdots,(p-1)\}$. Thus, we may choose the Mori cone
generators to be
\begin{equation}
    C_k=P_1\cdot I_k,\quad C_p=I_0\cdot I_1\, .
\end{equation}

\section{Linking forms}

\subsection{$\cC_\bR(Y^{p,q})$}
\label{app:L_{Y^{p,q}}}

The intersection form $Q_4$ ($Q_2=Q_4^T$) between $4$-cycles
($2$-cycles) and $2$-cycles ($4$-cycles) can be easily read from the
Mori cone generators to be
\begin{equation*}
    Q_4=
        \begin{pmatrix}
            -2&1&0&0&\cdots &0&0&0&0\\
            1&-2&1&0&\cdots&0&0&0&0\\
            \vdots\\
            0&0&0&0&\cdots&0&1&-2&1\\
            0&0&0&0&\cdots&0&0&1&-2\\
            -l&0&0&0&\cdots&0&0&0&0\\
          \end{pmatrix},
\end{equation*}
for even $l$, where $-2$ in the last row is in the column $l/2$ of $Q_4$ or in a more compact notation 
\begin{equation}
  \label{eq:Y^{p,q}-q-matrix}
  Q_4=(q_{i,j}), \quad 
  q_{i,j}=
  \begin{cases}
    \delta_{i, j-1}-2\delta_{i,j}+\delta_{i,j+1}, & \text{for}\quad i,j\in\{1,2,\cdots,p-1\},\\
    -l & \text{for} \quad i=p\; \text{and } j=1\\
    0, & \text{otherwise}.
  \end{cases}
\end{equation}

Now, we can calculate the homology groups of $Y^{p,q}$
using~\eqref{eq:homy}. We can easily determine the kernel and the
cokernel of $Q_4$ by finding the Smith normal form of $Q_4$,
which we call $S_4$. We find that $S_4$ has the form
\begin{equation}\label{s4}
  S_4 =  \begin{pmatrix}
    1&0&0&\cdots&0&0&0\\
    0&1&0&\cdots&0&0&0\\
    \vdots\\
    0&0&0&\cdots&0&1&0\\
    0&0&0&\cdots&0&0&\gcd(p,q)\\
    0&0&0&\cdots&0&0&0\\
  \end{pmatrix}
\end{equation}
or in more compact notation
\begin{equation}
  Q_4=(s_{i,j}), \quad s_{i,j}= 
            \begin{cases}
                \delta_{i, j}, & \text{for} \quad i,j\in\{1,2,\cdots,p-1\},\\
                \gcd(p,q)\delta_{i,p-1}\delta_{j,p-1}, & \text{for} \quad i=j=p-1,\\
                0, & \text{otherwise}.
            \end{cases}
\end{equation}

Hence, the image of $Q_4$ is
$\text{Im}(Q_4)=\mathbb{Z}^{p-2}+\gcd(p,q)\mathbb{Z}$ and its kernel
is zero. We have
\begin{equation*}
    \begin{split}
        &H_3(Y^{p,q})=\coker(Q_4)=\mathbb{Z}+\mathbb{Z}_{\gcd(p,q)},\\
        &H_4(Y^{p,q})=\ker(Q_4)=0\, .
    \end{split}
\end{equation*}
Similarly, since $Q_2=Q_4^T$ we find
\begin{equation*}
    H_2(Y^{p,q})=\text{ker}(Q_2)=\mathbb{Z}, \quad
    H_1(Y^{p,q})=\text{coker}(Q_2)=\mathbb{Z}_{\gcd(p,q)}\, .
\end{equation*}

\medskip

We now want to compute the linking pairing
\begin{equation}
  \sL_{Y^{p,q}}\colon \Tor H_{p-1}(Y^{p,q})\times \Tor H_{n-p-1}(Y^{p,q})\to \mathbb{Q}/\mathbb{Z}\, . 
\end{equation}
In our case the only homology groups with non-trivial torsion are
$H_3(Y^{p,q})$ and $H_1(Y^{p,q})$ so we may compute the linking of
$3$-cycles and $1$-cycles as follows. From~\eqref{link3}
\begin{equation}
    L(\partial\alpha^{'*}_i,\bar{\partial}{\beta^{'*}_j})=
    q^{-1}(\alpha^{'*}_i,{\beta^{'*}_j})=
    q^{-1}_{ij}
    \quad (\text{mod}\; 1)\, .
\end{equation}

We find
\begin{equation}\label{inverse}
    q^{-1}_{i,k}=
        \begin{cases}
            (i - j) + (p - j) c/2, & \text{for}\quad
            i \geq j\;,\; j < p \text{ and } i<p-1\\
            (p - j) c/2, & \text{for}\quad
            i < j\;,\; j < p \text{ and } i<p-1\\
            (p - 2 - j)/2 + (p - j) c/2, & \text{for}\quad j < p \text{ and } i = p - 1\\
            -i/l - pc/(2 l), & \text{for}\quad j = p  \text{ and } i < p - 1\\
            (1/l - p/(2 l)) - pc/(2 l), & \text{for}\quad j = p \text{ and } i = p - 1\; ,
        \end{cases}
\end{equation}
such that $q^{-1}q=I$. All that remains is to find the generators
$\alpha^{'*}_i$ and $\beta^{'*}_j$ defined above. This may be done by
tracking how the generators in the basis defined by the matrix
$q^{(T)}$ change as we switch basis by writing the matrix in its Smith
normal form $S^{(T)} $. Given the form of our matrix $q$ in
\eqref{eq:Y^{p,q}-q-matrix}), for $\beta^*_i$ and $\beta^{'*}_j$ the
generators of $\Hom (H_4(X_6), \mathbb{Z})$ in the $q$ basis and the $S$
basis, respectively, we find that, $\beta^{'*}_{p-1}=\beta^*_{p-1}$
where $\partial\beta^{'*}_{p-1}$ is the generator of $\Tor
H_{1}(Y)$. Similarly, for $\alpha^{*}_{\bar{i}}$ and
$\alpha^{'*}_{\bar{j}}$ the generators of $\Hom (H_2(X), \mathbb{Z})$
in the $q^T$ basis, and the $S^T$ basis, respectively, we find
$\alpha^{'*}_p=p'\alpha^*_1+q'\alpha^*_p$ such that,
$\partial\alpha^{'*}_p$ is the generator of $\Tor H_{3}(Y^{p,q})$ where,
$p'=\frac{p}{\text{gcd}(p,q)}$ and
$q'=\frac{lp-l}{\text{gcd}(p,q)}$. Therefore, the linking number is
\begin{equation}
    \begin{split}
    L(\partial\alpha^{'*}_p,\bar{\partial}{\beta^{'*}_{p-1}})&=
    L\big(p'\partial\alpha^*_1+q'\partial\alpha^*_p,\bar{\partial}{\beta^{*}_{p-1}}\big)\\&=
    p'q^{-1}_{1,p-1}+q'q^{-1}_{p,p-1}\\&=
    -\frac{1}{\text{gcd}(p,q)} 
    \quad (\text{mod}\; 1)\, ,
    \end{split}
\end{equation}
using (\ref{inverse}) and the bilinearity of the linking pairing.

\subsection{$\mathbb{C}^3/\mathbb{Z}_{2n+1}$}
\label{app:E0rankp}

Let $a=(-1,0)$, $b=(1,-1)$ and $I_i=(0,i)$ be the points on the toric
diagram with $i=0,1,2,\cdots,n$, and choose the triangulation such
that the $3$-dimensional cones are of the form $(a,b,I_0)$,
$(a,I_k,I_{k+1})$ and $(b,I_k,I_{k+1})$, where
$k=0,1,2,\cdots,n-1$. As before, from the toric diagram we have
\begin{equation}
    H_2(X_6)=H_4(X_6)=\bZ^n\, .
\end{equation}

\begin{table}[t]
\setlength{\tabcolsep}{4pt}
    \centering
    \begin{tabular}{ | m{3.5em} | m{.4cm}| m{.4cm} | m{.8cm} | m{.8cm} |m{0.4cm}| m{0.4cm}| m{0.7cm}| m{1.6cm}| m{1.4cm} |  m{0.4cm}| m{1.3cm}| m{1cm} | } \hline
    Curve&\centering $a$ &\centering$b$ &\centering $I_0$ &\centering $I_1$&\centering$I_2$ &\centering$\cdots$ &\centering $I_{k-1}$ &\centering $I_k$ &\centering $I_{k+1}$ &\centering $\cdots$ &\centering$I_{n-1}$ &$I_n$ \\ [0.5ex] 
    \hline
    \hline
    $a\cdot I_1$&\centering$0$ &\centering$0$&\centering$1$&\centering$-2$&\centering$1$&$\cdots$&\centering$0$&\centering$0$&\centering$0$&$\cdots$&\centering$0$&$0$\\ [0.5ex] 
    \hline
    $\cdots$ &&&&&&&&&&&& \\ [0.5ex] 
    \hline
    $a\cdot I_{k}$ &\centering$0$ &\centering $0$&\centering $0$ &\centering$0$&\centering $0$ &\centering$\cdots$ &\centering $1$ &\centering $-2$ &\centering $1$ &\centering$\cdots$ &\centering$0$&$0$\\[0.5ex]  
    \hline 
    $\cdots$ &&&&&&&&&&&& \\ [0.5ex] 
    \hline
    $I_0\cdot I_1$&\centering$1$ &\centering$1$&\centering$-3$&\centering$1$&\centering$0$&$\cdots$&\centering$0$&\centering$0$&\centering$0$&$\cdots$&\centering$0$&$0$\\ [0.5ex] 
    \hline
    $I_{k}\cdot I_{k+1}$ &\centering $1$ &\centering$1$ &\centering$0$ &\centering $0$ &\centering $0$&\centering$\cdots$ &\centering $0$ &\centering $-3-2k$ &\centering $2k+1$ &\centering $\cdots$ &\centering $0$ &$0$\\ [0.5ex] 
    \hline
    $\cdots$ &&&&&&&&&&&&  \\ [0.5ex] 
    \hline
    $I_{n-1}\cdot I_n$ &\centering$1$&\centering$1$&\centering$0$&\centering$0$&$\cdots$&\centering$0$&\centering$0$&\centering$0$&\centering $0$&$\cdots$&\centering$-2n-1$&$2n-1$\\ [0.5ex]
    \hline
\end{tabular}
\caption{The intersection numbers of the $(2P-1)$ compact curves
  $a\cdot I_k$, $I_k\cdot I_{k-1}$ and the $(n+3)$ points $a=(-1,0)$,
  $b=(1,-1)$ and $I_i=(0,i)$, where $i\in\{0,1,2,\cdots,n\}$ and
  $k\in\{0,1,2,\cdots,(n-1)\}$. We have omitted the result for the
  curves $b\cdot I_k$ and $a\cdot I_0$ as they give the same
  intersection numbers as $a\cdot I_k$ and $I_0\cdot I_1$ for each
  fixed $k$, respectively.}
\label{table:C3/Z3-curves}
\end{table}

From the intersection numbers given in table~\ref{table:C3/Z3-curves},
we deduce the equivalence relations (by subtracting the two relevant
rows in terms of $k$ for the latter relation)
\begin{equation}
    a\cdot I_i\equiv b\cdot I_i\, ,\quad a\cdot I_0\equiv I_0\cdot I_1\, ,\quad I_{k+1}\cdot I_{k+2}-I_{k}\cdot I_{k+1}\equiv (3+2k)a\cdot I_k\, .
\end{equation}
Therefore, we can choose the Mori cone generators $C_k$ to be the rows of table given by the intersection numbers for $a\cdot I_k$
\begin{equation}
    C_k= a\cdot I_k\, .
\end{equation}
The intersection form is 
\begin{equation}
    q_{i,j}=
        \begin{cases}

            \delta_{i,j}-2\delta_{i,j-1}+\delta_{i,j-2}\, ,& \text{for}\quad i\in\{1,2,..n-1\}\,,\, j\in\{1,2,..n\}\\
            -3\delta_{1,j}+\delta_{2,j}\, ,&\text{for}\quad i=n\, , j\in\{1,2,..n\}
        \end{cases}
\end{equation}
which has Smith normal form
\begin{equation}
    S_{i,j}=
        \begin{cases}
        \delta_{i,j}\, ,&\text{for}\quad i,j\in\{1,2,..n-1\}\\
        2n+1\, ,&\text{for}\quad i,j=n 
        \end{cases}
\end{equation}
From this we find
\begin{equation}
    \begin{split}
        &H_1(Y_5)=H_3(Y_5)=\coker(Q)=\mathbb{Z}_{2n+1},\\
        &H_2(Y_5)=H_4(Y_5)=\ker(Q)=0\, .
    \end{split}
\end{equation}
Now, to find the linking number, we track the effect on the generators
as we write $Q$ in its Smith normal form. We find $\alpha^*_n$ and
$\beta^*_n$ to be the generators of $\Hom (H_2(X), \mathbb{Z})$ and
$\Hom (H_4(X_6), \mathbb{Z})$, respectively such that,
$\partial\alpha^*_n$ and $\partial\beta^*_n$ are the generators of
$\Tor H_{3}(Y_5)$ and $\Tor H_{1}(Y_5)$, respectively. It can be shown
that the inverse of $q_{i,j}$ is
\begin{equation}
    -(2n+1) q_{i,j}^{-1}=
        \begin{cases}
            (2i-1)(n-j)\, ,& \text{for}\quad j<n\;\text{and}\; j+1\geq i\\
            (n-i+1)(2j+1)\, ,& \text{for}\quad  j+1\leq i\\
            (n-i+1)\, ,& \text{for}\quad  j=n
        \end{cases}
\end{equation}
i.e. $q^{-1}_{nn}=-\frac{1}{2n+1}$, and so we have 
\begin{equation}
  L(\partial\alpha^{*}_n,{\partial}{\beta^{*}_n})=
  q^{-1}_{nn}=-\frac{1}{2n+1}
  \mod 1\, .
\end{equation}

\bibliographystyle{JHEP}
\bibliography{refs}

\providecommand{\href}[2]{#2}\begingroup\raggedright\begin{thebibliography}{100}

\bibitem{DelZotto:2015isa}
M.~Del~Zotto, J.~J. Heckman, D.~S. Park and T.~Rudelius, \emph{{On the Defect
  Group of a 6D SCFT}},
  \href{http://dx.doi.org/10.1007/s11005-016-0839-5}{\emph{Lett.\ Math.\ Phys.}
  {\bf 106} (2016) 765--786}, [\href{http://arxiv.org/abs/1503.04806}{{\tt
  1503.04806}}].

\bibitem{Kapustin:2014gua}
A.~Kapustin and N.~Seiberg, \emph{{Coupling a QFT to a TQFT and Duality}},
  \href{http://dx.doi.org/10.1007/JHEP04(2014)001}{\emph{JHEP} {\bf 04} (2014)
  001}, [\href{http://arxiv.org/abs/1401.0740}{{\tt 1401.0740}}].

\bibitem{Gaiotto:2014kfa}
D.~Gaiotto, A.~Kapustin, N.~Seiberg and B.~Willett, \emph{{Generalized Global
  Symmetries}}, \href{http://dx.doi.org/10.1007/JHEP02(2015)172}{\emph{JHEP}
  {\bf 02} (2015) 172}, [\href{http://arxiv.org/abs/1412.5148}{{\tt
  1412.5148}}].

\bibitem{Sharpe:2015mja}
E.~Sharpe, \emph{{Notes on generalized global symmetries in QFT}},
  \href{http://dx.doi.org/10.1002/prop.201500048}{\emph{Fortsch. Phys.} {\bf
  63} (2015) 659--682}, [\href{http://arxiv.org/abs/1508.04770}{{\tt
  1508.04770}}].

\bibitem{Cordova:2018cvg}
C.~C{\'o}rdova, T.~T. Dumitrescu and K.~Intriligator, \emph{{Exploring 2-Group
  Global Symmetries}},
  \href{http://dx.doi.org/10.1007/JHEP02(2019)184}{\emph{JHEP} {\bf 02} (2019)
  184}, [\href{http://arxiv.org/abs/1802.04790}{{\tt 1802.04790}}].

\bibitem{Freed:2006ya}
D.~S. Freed, G.~W. Moore and G.~Segal, \emph{{The Uncertainty of Fluxes}},
  \href{http://dx.doi.org/10.1007/s00220-006-0181-3}{\emph{Commun. Math. Phys.}
  {\bf 271} (2007) 247--274}, [\href{http://arxiv.org/abs/hep-th/0605198}{{\tt
  hep-th/0605198}}].

\bibitem{Freed:2006yc}
D.~S. Freed, G.~W. Moore and G.~Segal, \emph{{Heisenberg Groups and
  Noncommutative Fluxes}},
  \href{http://dx.doi.org/10.1016/j.aop.2006.07.014}{\emph{Annals Phys.} {\bf
  322} (2007) 236--285}, [\href{http://arxiv.org/abs/hep-th/0605200}{{\tt
  hep-th/0605200}}].

\bibitem{Garcia-Etxebarria:2019cnb}
I.~García~Etxebarria, B.~Heidenreich and D.~Regalado, \emph{{IIB flux
  non-commutativity and the global structure of field theories}},
  \href{http://dx.doi.org/10.1007/JHEP10(2019)169}{\emph{JHEP} {\bf 10} (2019)
  169}, [\href{http://arxiv.org/abs/1908.08027}{{\tt 1908.08027}}].

\bibitem{Witten:2009at}
E.~Witten, \emph{{Geometric Langlands From Six Dimensions}},
  \href{http://arxiv.org/abs/0905.2720}{{\tt 0905.2720}}.

\bibitem{Heckman:2017uxe}
J.~J. Heckman and L.~Tizzano, \emph{{6D Fractional Quantum Hall Effect}},
  \href{http://dx.doi.org/10.1007/JHEP05(2018)120}{\emph{JHEP} {\bf 05} (2018)
  120}, [\href{http://arxiv.org/abs/1708.02250}{{\tt 1708.02250}}].

\bibitem{Gukov:2018iiq}
S.~Gukov, D.~Pei, P.~Putrov and C.~Vafa, \emph{{4-manifolds and topological
  modular forms}},  \href{http://arxiv.org/abs/1811.07884}{{\tt 1811.07884}}.

\bibitem{Eckhard:2019jgg}
J.~Eckhard, H.~Kim, S.~Schafer-Nameki and B.~Willett, \emph{{Higher-Form
  Symmetries, Bethe Vacua, and the 3d-3d Correspondence}},
  \href{http://dx.doi.org/10.1007/JHEP01(2020)101}{\emph{JHEP} {\bf 01} (2020)
  101}, [\href{http://arxiv.org/abs/1910.14086}{{\tt 1910.14086}}].

\bibitem{Dabholkar:2020fde}
A.~Dabholkar, P.~Putrov and E.~Witten, \emph{{Duality and Mock Modularity}},
  \href{http://arxiv.org/abs/2004.14387}{{\tt 2004.14387}}.

\bibitem{Cordova:2016emh}
C.~Cordova, T.~T. Dumitrescu and K.~Intriligator, \emph{{Multiplets of
  Superconformal Symmetry in Diverse Dimensions}},
  \href{http://dx.doi.org/10.1007/JHEP03(2019)163}{\emph{JHEP} {\bf 03} (2019)
  163}, [\href{http://arxiv.org/abs/1612.00809}{{\tt 1612.00809}}].

\bibitem{Chang:2018xmx}
C.-M. Chang, \emph{{5d and 6d SCFTs Have No Weak Coupling Limit}},
  \href{http://dx.doi.org/10.1007/JHEP09(2019)016}{\emph{JHEP} {\bf 09} (2019)
  016}, [\href{http://arxiv.org/abs/1810.04169}{{\tt 1810.04169}}].

\bibitem{Seiberg:1996bd}
N.~Seiberg, \emph{{Five-dimensional SUSY field theories, nontrivial fixed
  points and string dynamics}},
  \href{http://dx.doi.org/10.1016/S0370-2693(96)01215-4}{\emph{Phys. Lett.}
  {\bf B388} (1996) 753--760}, [\href{http://arxiv.org/abs/hep-th/9608111}{{\tt
  hep-th/9608111}}].

\bibitem{Morrison:1996xf}
D.~R. Morrison and N.~Seiberg, \emph{{Extremal transitions and five-dimensional
  supersymmetric field theories}},
  \href{http://dx.doi.org/10.1016/S0550-3213(96)00592-5}{\emph{Nucl. Phys.}
  {\bf B483} (1997) 229--247}, [\href{http://arxiv.org/abs/hep-th/9609070}{{\tt
  hep-th/9609070}}].

\bibitem{Intriligator:1997pq}
K.~A. Intriligator, D.~R. Morrison and N.~Seiberg, \emph{{Five-dimensional
  supersymmetric gauge theories and degenerations of Calabi-Yau spaces}},
  \href{http://dx.doi.org/10.1016/S0550-3213(97)00279-4}{\emph{Nucl. Phys.}
  {\bf B497} (1997) 56--100}, [\href{http://arxiv.org/abs/hep-th/9702198}{{\tt
  hep-th/9702198}}].

\bibitem{Aharony:1997ju}
O.~Aharony and A.~Hanany, \emph{{Branes, superpotentials and superconformal
  fixed points}},
  \href{http://dx.doi.org/10.1016/S0550-3213(97)00472-0}{\emph{Nucl. Phys.}
  {\bf B504} (1997) 239--271}, [\href{http://arxiv.org/abs/hep-th/9704170}{{\tt
  hep-th/9704170}}].

\bibitem{Leung:1997tw}
N.~C. Leung and C.~Vafa, \emph{{Branes and toric geometry}},
  \href{http://dx.doi.org/10.4310/ATMP.1998.v2.n1.a4}{\emph{Adv. Theor. Math.
  Phys.} {\bf 2} (1998) 91--118},
  [\href{http://arxiv.org/abs/hep-th/9711013}{{\tt hep-th/9711013}}].

\bibitem{Bergman:2013ala}
O.~Bergman, D.~Rodríguez-Gómez and G.~Zafrir, \emph{{Discrete $\theta$ and
  the 5d superconformal index}},
  \href{http://dx.doi.org/10.1007/JHEP01(2014)079}{\emph{JHEP} {\bf 01} (2014)
  079}, [\href{http://arxiv.org/abs/1310.2150}{{\tt 1310.2150}}].

\bibitem{Hayashi:2014hfa}
H.~Hayashi, Y.~Tachikawa and K.~Yonekura, \emph{{Mass-deformed T$_{N}$ as a
  linear quiver}}, \href{http://dx.doi.org/10.1007/JHEP02(2015)089}{\emph{JHEP}
  {\bf 02} (2015) 089}, [\href{http://arxiv.org/abs/1410.6868}{{\tt
  1410.6868}}].

\bibitem{Bergman:2014kza}
O.~Bergman and G.~Zafrir, \emph{{Lifting 4d dualities to 5d}},
  \href{http://dx.doi.org/10.1007/JHEP04(2015)141}{\emph{JHEP} {\bf 04} (2015)
  141}, [\href{http://arxiv.org/abs/1410.2806}{{\tt 1410.2806}}].

\bibitem{Zafrir:2014ywa}
G.~Zafrir, \emph{{Duality and enhancement of symmetry in 5d gauge theories}},
  \href{http://dx.doi.org/10.1007/JHEP12(2014)116}{\emph{JHEP} {\bf 12} (2014)
  116}, [\href{http://arxiv.org/abs/1408.4040}{{\tt 1408.4040}}].

\bibitem{DelZotto:2014hpa}
M.~Del~Zotto, J.~J. Heckman, A.~Tomasiello and C.~Vafa, \emph{{6d Conformal
  Matter}}, \href{http://dx.doi.org/10.1007/JHEP02(2015)054}{\emph{JHEP} {\bf
  02} (2015) 054}, [\href{http://arxiv.org/abs/1407.6359}{{\tt 1407.6359}}].

\bibitem{Bergman:2015dpa}
O.~Bergman and G.~Zafrir, \emph{{5d fixed points from brane webs and
  O7-planes}}, \href{http://dx.doi.org/10.1007/JHEP12(2015)163}{\emph{JHEP}
  {\bf 12} (2015) 163}, [\href{http://arxiv.org/abs/1507.03860}{{\tt
  1507.03860}}].

\bibitem{Zafrir:2015ftn}
G.~Zafrir, \emph{{Brane webs and $O5$-planes}},
  \href{http://dx.doi.org/10.1007/JHEP03(2016)109}{\emph{JHEP} {\bf 03} (2016)
  109}, [\href{http://arxiv.org/abs/1512.08114}{{\tt 1512.08114}}].

\bibitem{Kim:2016qqs}
H.-C. Kim, \emph{{Line defects and 5d instanton partition functions}},
  \href{http://dx.doi.org/10.1007/JHEP03(2016)199}{\emph{JHEP} {\bf 03} (2016)
  199}, [\href{http://arxiv.org/abs/1601.06841}{{\tt 1601.06841}}].

\bibitem{DelZotto:2015rca}
M.~Del~Zotto, C.~Vafa and D.~Xie, \emph{{Geometric engineering, mirror symmetry
  and $ 6{\mathrm{d}}_{\left(1,0\right)}\to
  4{\mathrm{d}}_{\left(\mathcal{N}=2\right)} $}},
  \href{http://dx.doi.org/10.1007/JHEP11(2015)123}{\emph{JHEP} {\bf 11} (2015)
  123}, [\href{http://arxiv.org/abs/1504.08348}{{\tt 1504.08348}}].

\bibitem{Hayashi:2015fsa}
H.~Hayashi, S.-S. Kim, K.~Lee, M.~Taki and F.~Yagi, \emph{{A new 5d description
  of 6d D-type minimal conformal matter}},
  \href{http://dx.doi.org/10.1007/JHEP08(2015)097}{\emph{JHEP} {\bf 08} (2015)
  097}, [\href{http://arxiv.org/abs/1505.04439}{{\tt 1505.04439}}].

\bibitem{Hayashi:2015vhy}
H.~Hayashi, S.-S. Kim, K.~Lee, M.~Taki and F.~Yagi, \emph{{More on 5d
  descriptions of 6d SCFTs}},
  \href{http://dx.doi.org/10.1007/JHEP10(2016)126}{\emph{JHEP} {\bf 10} (2016)
  126}, [\href{http://arxiv.org/abs/1512.08239}{{\tt 1512.08239}}].

\bibitem{Hayashi:2016abm}
H.~Hayashi, S.-S. Kim, K.~Lee and F.~Yagi, \emph{{Equivalence of several
  descriptions for 6d SCFT}},
  \href{http://dx.doi.org/10.1007/JHEP01(2017)093}{\emph{JHEP} {\bf 01} (2017)
  093}, [\href{http://arxiv.org/abs/1607.07786}{{\tt 1607.07786}}].

\bibitem{Xie:2017pfl}
D.~Xie and S.-T. Yau, \emph{{Three dimensional canonical singularity and five
  dimensional $ \mathcal{N} $ = 1 SCFT}},
  \href{http://dx.doi.org/10.1007/JHEP06(2017)134}{\emph{JHEP} {\bf 06} (2017)
  134}, [\href{http://arxiv.org/abs/1704.00799}{{\tt 1704.00799}}].

\bibitem{DelZotto:2017pti}
M.~Del~Zotto, J.~J. Heckman and D.~R. Morrison, \emph{{6D SCFTs and Phases of
  5D Theories}}, \href{http://dx.doi.org/10.1007/JHEP09(2017)147}{\emph{JHEP}
  {\bf 09} (2017) 147}, [\href{http://arxiv.org/abs/1703.02981}{{\tt
  1703.02981}}].

\bibitem{Alexandrov:2017mgi}
S.~Alexandrov, S.~Banerjee and P.~Longhi, \emph{{Rigid limit for
  hypermultiplets and five-dimensional gauge theories}},
  \href{http://dx.doi.org/10.1007/JHEP01(2018)156}{\emph{JHEP} {\bf 01} (2018)
  156}, [\href{http://arxiv.org/abs/1710.10665}{{\tt 1710.10665}}].

\bibitem{Ferlito:2017xdq}
G.~Ferlito, A.~Hanany, N.~Mekareeya and G.~Zafrir, \emph{{3d Coulomb branch and
  5d Higgs branch at infinite coupling}},
  \href{http://dx.doi.org/10.1007/JHEP07(2018)061}{\emph{JHEP} {\bf 07} (2018)
  061}, [\href{http://arxiv.org/abs/1712.06604}{{\tt 1712.06604}}].

\bibitem{Jefferson:2017ahm}
P.~Jefferson, H.-C. Kim, C.~Vafa and G.~Zafrir, \emph{{Towards Classification
  of 5d SCFTs: Single Gauge Node}},
  \href{http://arxiv.org/abs/1705.05836}{{\tt 1705.05836}}.

\bibitem{Jefferson:2018irk}
P.~Jefferson, S.~Katz, H.-C. Kim and C.~Vafa, \emph{{On Geometric
  Classification of 5d SCFTs}},
  \href{http://dx.doi.org/10.1007/JHEP04(2018)103}{\emph{JHEP} {\bf 04} (2018)
  103}, [\href{http://arxiv.org/abs/1801.04036}{{\tt 1801.04036}}].

\bibitem{Apruzzi:2018nre}
F.~Apruzzi, L.~Lin and C.~Mayrhofer, \emph{{Phases of 5d SCFTs from M-/F-theory
  on Non-Flat Fibrations}},  \href{http://arxiv.org/abs/1811.12400}{{\tt
  1811.12400}}.

\bibitem{Bhardwaj:2018yhy}
L.~Bhardwaj and P.~Jefferson, \emph{{Classifying 5d SCFTs via 6d SCFTs: Rank
  one}},  \href{http://arxiv.org/abs/1809.01650}{{\tt 1809.01650}}.

\bibitem{Closset:2018bjz}
C.~Closset, M.~Del~Zotto and V.~Saxena, \emph{{Five-dimensional SCFTs and gauge
  theory phases: an M-theory/type IIA perspective}},
  \href{http://dx.doi.org/10.21468/SciPostPhys.6.5.052}{\emph{SciPost Phys.}
  {\bf 6} (2019) 052}, [\href{http://arxiv.org/abs/1812.10451}{{\tt
  1812.10451}}].

\bibitem{DelZotto:2018tcj}
M.~Del~Zotto and G.~Lockhart, \emph{{Universal Features of BPS Strings in
  Six-dimensional SCFTs}},
  \href{http://dx.doi.org/10.1007/JHEP08(2018)173}{\emph{JHEP} {\bf 08} (2018)
  173}, [\href{http://arxiv.org/abs/1804.09694}{{\tt 1804.09694}}].

\bibitem{Bhardwaj:2018vuu}
L.~Bhardwaj and P.~Jefferson, \emph{{Classifying 5d SCFTs via 6d SCFTs:
  Arbitrary rank}},  \href{http://arxiv.org/abs/1811.10616}{{\tt 1811.10616}}.

\bibitem{Bhardwaj:2019hhd}
L.~Bhardwaj, \emph{{Revisiting the classifications of 6d SCFTs and LSTs}},
  \href{http://dx.doi.org/10.1007/JHEP03(2020)171}{\emph{JHEP} {\bf 03} (2020)
  171}, [\href{http://arxiv.org/abs/1903.10503}{{\tt 1903.10503}}].

\bibitem{Bhardwaj:2019ngx}
L.~Bhardwaj, \emph{{Dualities of 5d gauge theories from S-duality}},
  \href{http://arxiv.org/abs/1909.05250}{{\tt 1909.05250}}.

\bibitem{Bhardwaj:2019fzv}
L.~Bhardwaj, P.~Jefferson, H.-C. Kim, H.-C. Tarazi and C.~Vafa, \emph{{Twisted
  Circle Compactifications of 6d SCFTs}},
  \href{http://arxiv.org/abs/1909.11666}{{\tt 1909.11666}}.

\bibitem{Bhardwaj:2019xeg}
L.~Bhardwaj, \emph{{Do all $5d$ SCFTs descend from $6d$ SCFTs?}},
  \href{http://arxiv.org/abs/1912.00025}{{\tt 1912.00025}}.

\bibitem{Bhardwaj:2020gyu}
L.~Bhardwaj and G.~Zafrir, \emph{{Classification of 5d N=1 gauge theories}},
  \href{http://arxiv.org/abs/2003.04333}{{\tt 2003.04333}}.

\bibitem{Bhardwaj:2020kim}
L.~Bhardwaj, \emph{{More 5d KK theories}},
  \href{http://arxiv.org/abs/2005.01722}{{\tt 2005.01722}}.

\bibitem{Apruzzi:2019kgb}
F.~Apruzzi, S.~Schaefer-Nameki and Y.-N. Wang, \emph{{5d SCFTs from Decoupling
  and Gluing}},  \href{http://arxiv.org/abs/1912.04264}{{\tt 1912.04264}}.

\bibitem{Apruzzi:2019enx}
F.~Apruzzi, C.~Lawrie, L.~Lin, S.~Schaefer-Nameki and Y.-N. Wang, \emph{{Fibers
  add Flavor, Part II: 5d SCFTs, Gauge Theories, and Dualities}},
  \href{http://dx.doi.org/10.1007/JHEP03(2020)052}{\emph{JHEP} {\bf 03} (2020)
  052}, [\href{http://arxiv.org/abs/1909.09128}{{\tt 1909.09128}}].

\bibitem{Apruzzi:2019vpe}
F.~Apruzzi, C.~Lawrie, L.~Lin, S.~Schaefer-Nameki and Y.-N. Wang, \emph{{5d
  Superconformal Field Theories and Graphs}},
  \href{http://dx.doi.org/10.1016/j.physletb.2019.135077}{\emph{Phys. Lett. B}
  {\bf 800} (2020) 135077}, [\href{http://arxiv.org/abs/1906.11820}{{\tt
  1906.11820}}].

\bibitem{Apruzzi:2019opn}
F.~Apruzzi, C.~Lawrie, L.~Lin, S.~Schaefer-Nameki and Y.-N. Wang, \emph{{Fibers
  add Flavor, Part I: Classification of 5d SCFTs, Flavor Symmetries and BPS
  States}}, \href{http://dx.doi.org/10.1007/JHEP11(2019)068}{\emph{JHEP} {\bf
  11} (2019) 068}, [\href{http://arxiv.org/abs/1907.05404}{{\tt 1907.05404}}].

\bibitem{Closset:2019juk}
C.~Closset and M.~Del~Zotto, \emph{{On 5d SCFTs and their BPS quivers. Part I:
  B-branes and brane tilings}},  \href{http://arxiv.org/abs/1912.13502}{{\tt
  1912.13502}}.

\bibitem{Kim:2019dqn}
H.-C. Kim, S.-S. Kim and K.~Lee, \emph{{Higgsing and Twisting of 6d $D_N$ gauge
  theories}},  \href{http://arxiv.org/abs/1908.04704}{{\tt 1908.04704}}.

\bibitem{BenettiGenolini:2019zth}
P.~Benetti~Genolini, M.~Honda, H.-C. Kim, D.~Tong and C.~Vafa, \emph{{Evidence
  for a Non-Supersymmetric 5d CFT from Deformations of 5d $SU(2)$ SYM}},
  \href{http://dx.doi.org/10.1007/JHEP05(2020)058}{\emph{JHEP} {\bf 05} (2020)
  058}, [\href{http://arxiv.org/abs/2001.00023}{{\tt 2001.00023}}].

\bibitem{Hayashi:2019fsa}
H.~Hayashi, P.~Jefferson, H.-C. Kim, K.~Ohmori and C.~Vafa, \emph{{SCFTs,
  Holography, and Topological Strings}},
  \href{http://arxiv.org/abs/1905.00116}{{\tt 1905.00116}}.

\bibitem{Bourget:2020gzi}
A.~Bourget, J.~F. Grimminger, A.~Hanany, M.~Sperling and Z.~Zhong,
  \emph{{Magnetic Quivers from Brane Webs with O5 Planes}},
  \href{http://arxiv.org/abs/2004.04082}{{\tt 2004.04082}}.

\bibitem{Bourget:2019rtl}
A.~Bourget, S.~Cabrera, J.~F. Grimminger, A.~Hanany and Z.~Zhong, \emph{{Brane
  Webs and Magnetic Quivers for SQCD}},
  \href{http://dx.doi.org/10.1007/JHEP03(2020)176}{\emph{JHEP} {\bf 03} (2020)
  176}, [\href{http://arxiv.org/abs/1909.00667}{{\tt 1909.00667}}].

\bibitem{Cabrera:2019izd}
S.~Cabrera, A.~Hanany and M.~Sperling, \emph{{Magnetic quivers, Higgs branches,
  and 6d $N$=(1,0) theories}},
  \href{http://dx.doi.org/10.1007/JHEP06(2019)071}{\emph{JHEP} {\bf 06} (2019)
  071}, [\href{http://arxiv.org/abs/1904.12293}{{\tt 1904.12293}}].

\bibitem{Cabrera:2018jxt}
S.~Cabrera, A.~Hanany and F.~Yagi, \emph{{Tropical Geometry and Five
  Dimensional Higgs Branches at Infinite Coupling}},
  \href{http://dx.doi.org/10.1007/JHEP01(2019)068}{\emph{JHEP} {\bf 01} (2019)
  068}, [\href{http://arxiv.org/abs/1810.01379}{{\tt 1810.01379}}].

\bibitem{Hayashi:2019jvx}
H.~Hayashi, S.-S. Kim, K.~Lee and F.~Yagi, \emph{{Complete prepotential for 5d
  $ \mathcal{N} $ = 1 superconformal field theories}},
  \href{http://dx.doi.org/10.1007/JHEP02(2020)074}{\emph{JHEP} {\bf 02} (2020)
  074}, [\href{http://arxiv.org/abs/1912.10301}{{\tt 1912.10301}}].

\bibitem{Hayashi:2019yxj}
H.~Hayashi, S.-S. Kim, K.~Lee and F.~Yagi, \emph{{Rank-3 antisymmetric matter
  on 5-brane webs}},
  \href{http://dx.doi.org/10.1007/JHEP05(2019)133}{\emph{JHEP} {\bf 05} (2019)
  133}, [\href{http://arxiv.org/abs/1902.04754}{{\tt 1902.04754}}].

\bibitem{Hayashi:2018lyv}
H.~Hayashi, S.-S. Kim, K.~Lee and F.~Yagi, \emph{{Dualities and 5-brane webs
  for 5d rank 2 SCFTs}},
  \href{http://dx.doi.org/10.1007/JHEP12(2018)016}{\emph{JHEP} {\bf 12} (2018)
  016}, [\href{http://arxiv.org/abs/1806.10569}{{\tt 1806.10569}}].

\bibitem{Hayashi:2017jze}
H.~Hayashi and K.~Ohmori, \emph{{5d/6d DE instantons from trivalent gluing of
  web diagrams}}, \href{http://dx.doi.org/10.1007/JHEP06(2017)078}{\emph{JHEP}
  {\bf 06} (2017) 078}, [\href{http://arxiv.org/abs/1702.07263}{{\tt
  1702.07263}}].

\bibitem{Hayashi:2015zka}
H.~Hayashi, S.-S. Kim, K.~Lee and F.~Yagi, \emph{{6d SCFTs, 5d Dualities and
  Tao Web Diagrams}},
  \href{http://dx.doi.org/10.1007/JHEP05(2019)203}{\emph{JHEP} {\bf 05} (2019)
  203}, [\href{http://arxiv.org/abs/1509.03300}{{\tt 1509.03300}}].

\bibitem{Saxena:2019wuy}
V.~Saxena, \emph{{Rank-two 5d SCFTs from M-theory at isolated toric
  singularities: a systematic study}},
  \href{http://dx.doi.org/10.1007/JHEP04(2020)198}{\emph{JHEP} {\bf 20} (2020)
  198}, [\href{http://arxiv.org/abs/1911.09574}{{\tt 1911.09574}}].

\bibitem{Garozzo:2020pmz}
I.~Garozzo, N.~Mekareeya, M.~Sacchi and G.~Zafrir, \emph{{Symmetry enhancement
  and duality walls in 5d gauge theories}},
  \href{http://arxiv.org/abs/2003.07373}{{\tt 2003.07373}}.

\bibitem{Tachikawa:2013hya}
Y.~Tachikawa, \emph{{On the 6d origin of discrete additional data of 4d gauge
  theories}}, \href{http://dx.doi.org/10.1007/JHEP05(2014)020}{\emph{JHEP} {\bf
  05} (2014) 020}, [\href{http://arxiv.org/abs/1309.0697}{{\tt 1309.0697}}].

\bibitem{Morrison:2020ool}
D.~R. Morrison, S.~Schafer-Nameki and B.~Willett, \emph{{Higher-Form Symmetries
  in 5d}},  \href{http://arxiv.org/abs/2005.12296}{{\tt 2005.12296}}.

\bibitem{deBoer:2001wca}
J.~de~Boer, R.~Dijkgraaf, K.~Hori, A.~Keurentjes, J.~Morgan, D.~R. Morrison
  et~al., \emph{{Triples, fluxes, and strings}}, {\emph{Adv. Theor. Math.
  Phys.} {\bf 4} (2002) 995--1186},
  [\href{http://arxiv.org/abs/hep-th/0103170}{{\tt hep-th/0103170}}].

\bibitem{Minahan:2015jta}
J.~A. Minahan and M.~Zabzine, \emph{{Gauge theories with 16 supersymmetries on
  spheres}}, \href{http://dx.doi.org/10.1007/JHEP03(2015)155}{\emph{JHEP} {\bf
  03} (2015) 155}, [\href{http://arxiv.org/abs/1502.07154}{{\tt 1502.07154}}].

\bibitem{Polydorou:2017jha}
K.~Polydorou, A.~Roc{\'e}n and M.~Zabzine, \emph{{7D supersymmetric Yang-Mills
  on curved manifolds}},
  \href{http://dx.doi.org/10.1007/JHEP12(2017)152}{\emph{JHEP} {\bf 12} (2017)
  152}, [\href{http://arxiv.org/abs/1710.09653}{{\tt 1710.09653}}].

\bibitem{Iakovidis:2020znp}
N.~Iakovidis, J.~Qiu, A.~Roc{\'e}n and M.~Zabzine, \emph{{7D supersymmetric
  Yang-Mills on hypertoric 3-Sasakian manifolds}},
  \href{http://arxiv.org/abs/2003.12461}{{\tt 2003.12461}}.

\bibitem{Gaiotto:2010be}
D.~Gaiotto, G.~W. Moore and A.~Neitzke, \emph{{Framed BPS States}},
  \href{http://dx.doi.org/10.4310/ATMP.2013.v17.n2.a1}{\emph{Adv. Theor. Math.
  Phys.} {\bf 17} (2013) 241--397}, [\href{http://arxiv.org/abs/1006.0146}{{\tt
  1006.0146}}].

\bibitem{Aharony:2013hda}
O.~Aharony, N.~Seiberg and Y.~Tachikawa, \emph{{Reading between the lines of
  four-dimensional gauge theories}},
  \href{http://dx.doi.org/10.1007/JHEP08(2013)115}{\emph{JHEP} {\bf 08} (2013)
  115}, [\href{http://arxiv.org/abs/1305.0318}{{\tt 1305.0318}}].

\bibitem{Aharony:1998qu}
O.~Aharony and E.~Witten, \emph{{Anti-de Sitter space and the center of the
  gauge group}},
  \href{http://dx.doi.org/10.1088/1126-6708/1998/11/018}{\emph{JHEP} {\bf 11}
  (1998) 018}, [\href{http://arxiv.org/abs/hep-th/9807205}{{\tt
  hep-th/9807205}}].

\bibitem{Witten:1998wy}
E.~Witten, \emph{{AdS / CFT correspondence and topological field theory}},
  \href{http://dx.doi.org/10.1088/1126-6708/1998/12/012}{\emph{JHEP} {\bf 12}
  (1998) 012}, [\href{http://arxiv.org/abs/hep-th/9812012}{{\tt
  hep-th/9812012}}].

\bibitem{Witten:1996md}
E.~Witten, \emph{{On flux quantization in M theory and the effective action}},
  \href{http://dx.doi.org/10.1016/S0393-0440(96)00042-3}{\emph{J. Geom. Phys.}
  {\bf 22} (1997) 1--13}, [\href{http://arxiv.org/abs/hep-th/9609122}{{\tt
  hep-th/9609122}}].

\bibitem{Sati:2013rxa}
H.~Sati, \emph{{Framed M-branes, corners, and topological invariants}},
  \href{http://dx.doi.org/10.1063/1.5007185}{\emph{J. Math. Phys.} {\bf 59}
  (2018) 062304}, [\href{http://arxiv.org/abs/1310.1060}{{\tt 1310.1060}}].

\bibitem{Hikami:2009ze}
K.~Hikami, \emph{{Decomposition of Witten-Reshetikhin-Turaev invariant: Linking
  pairing and modular forms}}, {\emph{AMS/IP Stud. Adv. Math.} {\bf 50} (2011)
  131--151}.

\bibitem{Mumford:1338263}
D.~Mumford, \emph{{Tata Lectures on Theta, 1}}.
\newblock Modern Birk{\"a}user Classics. Springer, Dordrecht, 2007.

\bibitem{Garcia-Etxebarria:2018ajm}
I.~García-Etxebarria and M.~Montero, \emph{{Dai-Freed anomalies in particle
  physics}}, \href{http://dx.doi.org/10.1007/JHEP08(2019)003}{\emph{JHEP} {\bf
  08} (2019) 003}, [\href{http://arxiv.org/abs/1808.00009}{{\tt 1808.00009}}].

\bibitem{Witten:1985bt}
E.~Witten, \emph{{Topological Tools in Ten-dimensional Physics}},
  \href{http://dx.doi.org/10.1142/S0217751X86000034}{\emph{Int. J. Mod. Phys.}
  {\bf A1} (1986) 39}.

\bibitem{Hatcher}
A.~Hatcher, \emph{Algebraic Topology}.
\newblock Algebraic Topology. Cambridge University Press, 2002.

\bibitem{Sharpe:2019ddn}
E.~Sharpe, \emph{{Undoing decomposition}},
  \href{http://dx.doi.org/10.1142/S0217751X19502336}{\emph{Int. J. Mod. Phys.
  A} {\bf 34} (2020) 1950233}, [\href{http://arxiv.org/abs/1911.05080}{{\tt
  1911.05080}}].

\bibitem{Tanizaki:2019rbk}
Y.~Tanizaki and M.~{\"U}nsal, \emph{{Modified instanton sum in QCD and
  higher-groups}}, \href{http://dx.doi.org/10.1007/JHEP03(2020)123}{\emph{JHEP}
  {\bf 03} (2020) 123}, [\href{http://arxiv.org/abs/1912.01033}{{\tt
  1912.01033}}].

\bibitem{Gu:2020ivl}
W.~Gu, E.~Sharpe and H.~Zou, \emph{{Notes on two-dimensional pure
  supersymmetric gauge theories}},  \href{http://arxiv.org/abs/2005.10845}{{\tt
  2005.10845}}.

\bibitem{BF}
G.~Barthel and K.~H. Fieseler, \emph{Invariant divisors and homology of compact
  complex toric varieties},
  \href{http://dx.doi.org/10.1007/BF02362565}{\emph{Journal of Mathematical
  Sciences} {\bf 82} (Dec, 1996) 3615--3624}.

\bibitem{Cox}
D.~Cox, J.~Little and H.~Schenck, \emph{Toric Varieties}.
\newblock Graduate studies in mathematics. American Mathematical Society, 2011.

\bibitem{Dan}
V.~Danilov, \emph{The geometry of toric varieties},
  \href{http://dx.doi.org/10.1070/RM1978v033n02ABEH002305}{\emph{Russian
  Mathematical Surveys - RUSS MATH SURVEY-ENGL TR} {\bf 33} (04, 1978)
  97--154}.

\bibitem{FS}
S.~Friedl, \emph{{Algebraic Topology I - V}},  2020.

\bibitem{CLS}
D.~A. Cox, J.~B. Little and H.~K. Schenck, \emph{Toric Varieties}.
\newblock Graduate Studies in Mathematics. AMS, 2011.

\bibitem{Ref}
S.~Reffert, \emph{{The Geometer's Toolkit to String Compactifications}},  in
  \emph{{Conference on String and M Theory Approaches to Particle Physics and
  Cosmology Florence, Italy, June 13-15, 2007}}, 2007.
\newblock \href{http://arxiv.org/abs/0706.1310}{{\tt 0706.1310}}.

\bibitem{Denef:2008wq}
F.~Denef, \emph{{Les Houches Lectures on Constructing String Vacua}},
  {\emph{Les Houches} {\bf 87} (2008) 483--610},
  [\href{http://arxiv.org/abs/0803.1194}{{\tt 0803.1194}}].

\bibitem{Hori:2003ic}
K.~Hori, S.~Katz, A.~Klemm, R.~Pandharipande, R.~Thomas, C.~Vafa et~al.,
  \emph{{Mirror symmetry}}, vol.~1 of \emph{Clay mathematics monographs}.
\newblock AMS, Providence, USA, 2003.

\bibitem{He17}
Y.-H. He, R.-K. Seong and S.-T. Yau, \emph{{Calabi–Yau Volumes and Reflexive
  Polytopes}}, \href{http://dx.doi.org/10.1007/s00220-018-3128-6}{\emph{Commun.
  Math. Phys.} {\bf 361} (2018) 155--204},
  [\href{http://arxiv.org/abs/1704.03462}{{\tt 1704.03462}}].

\bibitem{sage}
W.~Stein et~al., \emph{{S}age {M}athematics {S}oftware ({V}ersion 9.1)}.
\newblock The Sage Development Team, 2020.

\bibitem{Garcia-Etxebarria:2016bpb}
I.~García-Etxebarria and B.~Heidenreich, \emph{{S-duality in $\mathscr{N} =$ 1
  orientifold SCFTs}},
  \href{http://dx.doi.org/10.1002/prop.201700013}{\emph{Fortsch. Phys.} {\bf
  65} (2017) 1700013}, [\href{http://arxiv.org/abs/1612.00853}{{\tt
  1612.00853}}].

\bibitem{Gauntlett:2004yd}
J.~P. Gauntlett, D.~Martelli, J.~Sparks and D.~Waldram, \emph{{Sasaki-Einstein
  metrics on $S^2 \times S^3$}},
  \href{http://dx.doi.org/10.4310/ATMP.2004.v8.n4.a3}{\emph{Adv. Theor. Math.
  Phys.} {\bf 8} (2004) 711--734},
  [\href{http://arxiv.org/abs/hep-th/0403002}{{\tt hep-th/0403002}}].

\bibitem{Aharony:1997bh}
O.~Aharony, A.~Hanany and B.~Kol, \emph{{Webs of (p,q) five-branes,
  five-dimensional field theories and grid diagrams}},
  \href{http://dx.doi.org/10.1088/1126-6708/1998/01/002}{\emph{JHEP} {\bf 01}
  (1998) 002}, [\href{http://arxiv.org/abs/hep-th/9710116}{{\tt
  hep-th/9710116}}].

\bibitem{Argyres:1995jj}
P.~C. Argyres and M.~R. Douglas, \emph{{New phenomena in SU(3) supersymmetric
  gauge theory}},
  \href{http://dx.doi.org/10.1016/0550-3213(95)00281-V}{\emph{Nucl. Phys. B}
  {\bf 448} (1995) 93--126}, [\href{http://arxiv.org/abs/hep-th/9505062}{{\tt
  hep-th/9505062}}].

\bibitem{Benini:2009gi}
F.~Benini, S.~Benvenuti and Y.~Tachikawa, \emph{{Webs of five-branes and N=2
  superconformal field theories}},
  \href{http://dx.doi.org/10.1088/1126-6708/2009/09/052}{\emph{JHEP} {\bf 09}
  (2009) 052}, [\href{http://arxiv.org/abs/0906.0359}{{\tt 0906.0359}}].

\bibitem{Eckhard:2020jyr}
J.~Eckhard, S.~Schafer-Nameki and Y.-N. Wang, \emph{{Trifectas for $T_N$ in
  5d}},  \href{http://arxiv.org/abs/2004.15007}{{\tt 2004.15007}}.

\bibitem{10.2307/24906437}
L.~Foscolo and M.~Haskins, \emph{New $g_2$-holonomy cones and exotic nearly
  k{\"a}hler structures on $s^6$ and $s^3 \times s^3$}, {\emph{Annals of
  Mathematics} {\bf 185} (2017) 59--130}.

\bibitem{Acharya:2004qe}
B.~S. Acharya and S.~Gukov, \emph{{M theory and singularities of exceptional
  holonomy manifolds}},
  \href{http://dx.doi.org/10.1016/j.physrep.2003.10.017}{\emph{Phys. Rept.}
  {\bf 392} (2004) 121--189}, [\href{http://arxiv.org/abs/hep-th/0409191}{{\tt
  hep-th/0409191}}].

\bibitem{tHooft:1977nqb}
G.~'t~Hooft, \emph{{On the Phase Transition Towards Permanent Quark
  Confinement}},
  \href{http://dx.doi.org/10.1016/0550-3213(78)90153-0}{\emph{Nucl. Phys. B}
  {\bf 138} (1978) 1--25}.

\bibitem{bryant1989}
R.~L. Bryant and S.~M. Salamon, \emph{On the construction of some complete
  metrics with exceptional holonomy},
  \href{http://dx.doi.org/10.1215/S0012-7094-89-05839-0}{\emph{Duke Math. J.}
  {\bf 58} (06, 1989) 829--850}.

\bibitem{Gibbons:1989er}
G.~Gibbons, D.~N. Page and C.~Pope, \emph{{Einstein Metrics on S**3 R**3 and
  R**4 Bundles}}, \href{http://dx.doi.org/10.1007/BF02104500}{\emph{Commun.
  Math. Phys.} {\bf 127} (1990) 529}.

\bibitem{Atiyah:2000zz}
M.~Atiyah, J.~M. Maldacena and C.~Vafa, \emph{{An M theory flop as a large N
  duality}}, \href{http://dx.doi.org/10.1063/1.1376159}{\emph{J. Math. Phys.}
  {\bf 42} (2001) 3209--3220}, [\href{http://arxiv.org/abs/hep-th/0011256}{{\tt
  hep-th/0011256}}].

\bibitem{Acharya:2001hq}
B.~S. Acharya, \emph{{Confining strings from G(2) holonomy space-times}},
  \href{http://arxiv.org/abs/hep-th/0101206}{{\tt hep-th/0101206}}.

\bibitem{Acharya:2001dz}
B.~S. Acharya and C.~Vafa, \emph{{On domain walls of N=1 supersymmetric
  Yang-Mills in four-dimensions}},
  \href{http://arxiv.org/abs/hep-th/0103011}{{\tt hep-th/0103011}}.

\end{thebibliography}\endgroup

\end{document}